%% file: usenix.tex
\documentclass[letterpaper,10pt, top=1in, bottom=1in, left=1.25in, right=1.25in]{article}
\usepackage{fullpage}
\usepackage{usenix,epsfig,endnotes}
\usepackage{caption}
\usepackage{subcaption}
\usepackage{tcolorbox}
\usepackage{pifont}

\usepackage{multirow}
\usepackage{graphicx}
\usepackage{float} 
\usepackage{longtable}
\usepackage{caption} 
\usepackage{array} 
\usepackage{tabularx}

\usepackage{setspace}
\begin{document}

\newcommand{\fixme}[1]{{\color{red} #1}}
\newcommand{\tanusree}[1]{{\color{blue}\textbf{(Tanusree: #1)}}}
\newcommand{\ayae}[1]{{\color{orange}\textbf{(Ayae: #1)}}}
\newcommand{\sandhi}[1]{{\color{cyan}\textbf{(Sandhi: #1)}}}

\date{}
\title{
Personhood Credentials: 
Human-Centered Design Recommendation Balancing Security, Usability, and Trust}


\author{
{\rm Ayae Ide\textsuperscript{1}, Tanusree Sharma\textsuperscript{1}}\\
{\rm \textsuperscript{1}Pennsylvania State University}\\
{\rm \small \{ayaeide, tanusree.sharma\}@psu.edu}
}

\maketitle
\vspace{-10mm}
\begin{figure*}[!h]
	\centering
	\includegraphics[width=\linewidth]{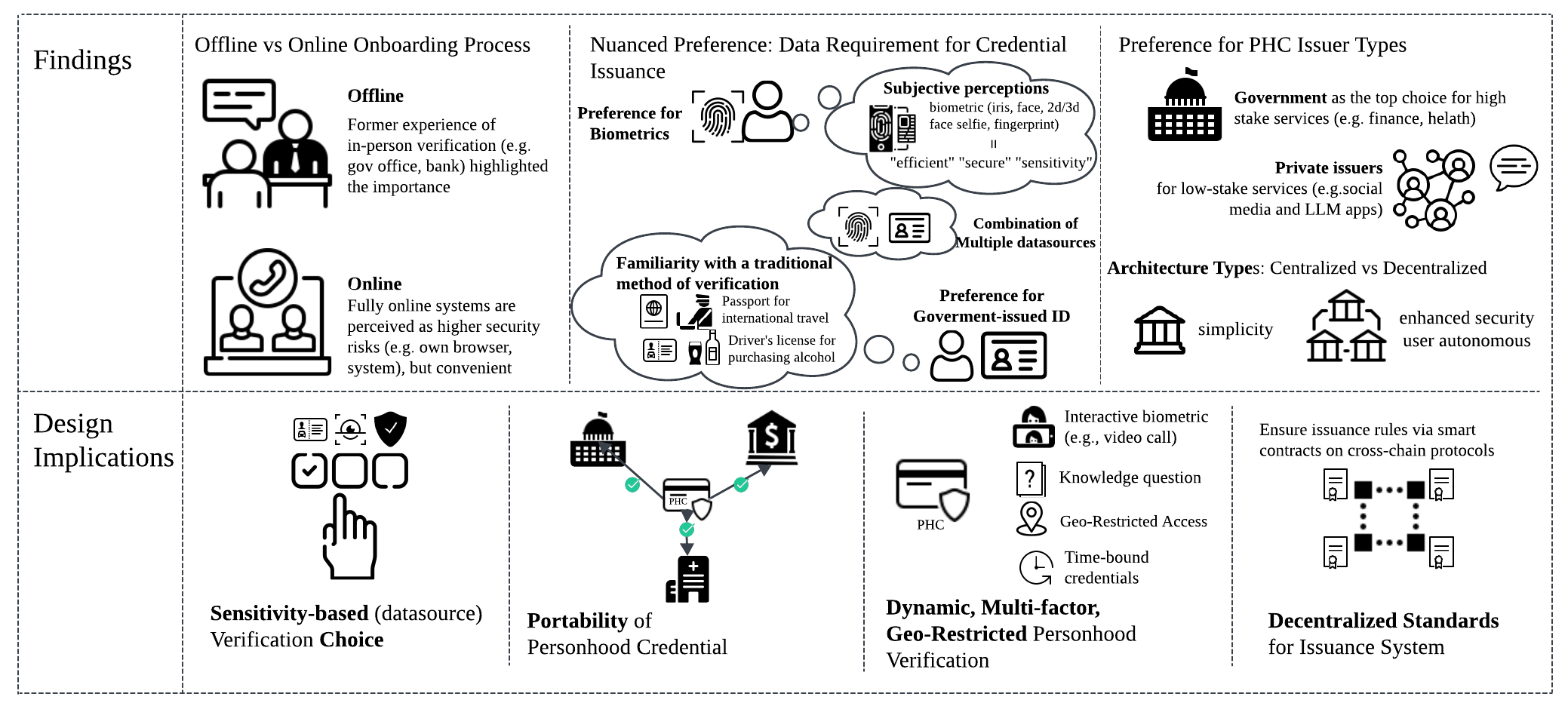}

	\caption{Overview of key findings and design implications from user interviews. Our study examines user preferences for onboarding processes (offline vs. online), credential types (e.g., biometrics, government-issued IDs), and issuer types (government vs. private companies). Additionally, we propose design implications, including verification choice facilitation, credential portability, dynamic multi-factor verification, and decentralized issuance standards.}
\label{fig:teaser}
\end{figure*}
\vspace{-5mm}

\subsection*{Abstract}
Building on related concepts, like, decentralized identifiers (DIDs), proof-of-personhood, anonymous credentials, personhood credentials (PHCs) emerged as an alternative approach, enabling individuals to verify to digital service providers that they are a person without disclosing additional information.
However, new technologies might introduce some friction due to users’ misunderstandings and mismatched expectations. Despite their growing importance, limited research has been done on users’ perceptions and preferences regarding PHCs. To address this gap, we conducted competitive analysis, and semi-structured online user interviews (N=23) to provide concrete design recommendations for PHCs that incorporate user needs, adoption rules, and preferences. Our study (i) surfaces how people reason about unknown privacy and security guarantees of PHCs compared to current verification methods; (ii) presents the impact of several factors on how people would like to onboard and manage PHCs, including, trusted issuers (e.g. gov), ground truth data to issue PHC (e.g biometrics, physical id), and issuance system (e.g. centralized vs decentralized). In a think-aloud conceptual design session, participants recommended conceptualized design, such as periodic biometrics verification, time-bound credentials, visually interactive human-check, and supervision of government for issuance system. We propose actionable designs reflecting users’ preferences.
\input{Sections/1_Introduction}
\input{Sections/2_Related_work}

\input{Sections/3_Method}

\input{Sections/Results_RQ1}

\input{Sections/Results_RQ2}

\input{Sections/4_Discussion}

{\footnotesize \bibliographystyle{acm}
\bibliography{sample-base}}

\appendix
\input{Sections/Appendix}


\end{document}

%% file: Sections/1_Introduction.tex
\vspace{-2mm}
\section{Introduction} 
\vspace{-2mm}
\label{sec:introduction}
Identity management has long been a cornerstone of user-facing systems such as social media platforms, gaming environments, and collaborative tools, and has increasingly become integral to Human-AI systems~\cite{gorwa2020unpacking, cetinkaya2007verification}. The recent proliferation of artificial intelligence (AI) has rendered traditional methods of multi-factor~\cite{aloul2009two}, and CAPTCHA~\cite{von-Ahn2003-wr} verification unreliable, as AI can now generate highly convincing fake human interactions~\cite{carlini2024poisoning, akhtar2024sok, goodrich2023battling}. 
Amid growing concerns over how major technology companies handle user data
~\cite{nytimesCambridgeAnalytica}, there has been a noticeable technological and ideological shift toward decentralized identity systems, commonly known as self-sovereign identity~\cite{mahula2021blockchain}. 

One widely recognized approach involves decentralized identifiers (DIDs). Emerging proposed systems, DECO~\cite{zhang2020deco}, Town-Crier~\cite{zhang2016town}  exemplify this model where users authorize the release of personal credentials from user devices to websites for proving certain characteristics about themselves. Although initiatives like the W3C Decentralized Identifier Working Group seek to establish standards for decentralized identity~\cite{identityDecentralizedIdentity, w3cccgDecentralizedIdentifiers}, many proposed frameworks struggle to meet both technical and usability requirements. Recent efforts, such as CanDID, have made progress in areas like usable key recovery~\cite{maram2021candid}.
Building on concepts, such as, decentralized identifiers (DIDs), proof-of-personhood \cite{ford2020identity}, anonymous credentials, and Personhood credentials (PHCs) have emerged as an alternative approach. PHCs allow individuals to verify their personhood to digital service providers without revealing additional personal information~\cite{adler2024personhood} and verified through zero-knowledge proofs. PHCs are designed to ensure that each credential corresponds to a unique, real individual~\cite{ford2020identity}. For example, Worldcoin has implemented a PHC-based identity system called World ID~\cite{de2024personhood}, which uses an "Orb" iris scanner for personhood verification.

Despite their growing importance
of personhood credentials and similar tools, there is a lack of understanding about how PHCs could be designed
from a user-centered perspective, and in particular what factors might influence users' preferences with regard to onboard and manage PHCs. New technologies might introduce some friction due to users’ misunderstandings and mismatched expectations. Consequently, to address this gap and to
complement existing system-centered approaches, our paper
explores a user-centered approach. 

Towards this goal, 
We began with a formative analysis, including a competitive review of existing personhood verification and related systems. This involved examining user reviews from app stores, analyzing white papers and analysing UI/UX of available system or proof of concepts.
to identify current challenges. 
Subsequently, we conducted interviews to explore user preferences and the factors (RQ1) influencing their choices of personhood credentials (RQ2). Finally, we conceptualized participants' desired approaches in managing personhood credentials (RQ3)\\

\vspace{-4pt}
\begin{tcolorbox}
\textbf{RQ1}: What are the users' perceptions of personhood credentials to verify themselves as legitimate and unique individuals in online interactions?\\
\textbf{RQ2}:  What factors influence user preferences 
on how people would like to verify themselves as legitimate and unique individuals in online interactions?\\
\textbf{RQ3}: How can personhood credentials be designed to ensure usability and security, enabling users to verify themselves as legitimate and unique individuals in online interactions?
\end{tcolorbox}
\vspace{5pt}
\textbf{Findings} Our study uncovered a wide range of users' perceptions regarding PHCs. 
Participants exhibited skepticism towards PHCs, partly because of \textit{``unknown risk''} vectors as a new technology compared to traditional verification. 
Despite these concerns, we find diverse levels of adoption preferences influenced by the \textit{``type of data required''} for PHC credential issuance and verification as well as personal \textit{``security standards''} for different services (e.g, finance, health, government related). 
Our findings also indicate benefits including aspects that promote fairness by ensuring opportunities for legitimate users in platforms such, as gaming, and survey platforms.
We surfaced nuanced preferences of how participants would like to onboard and manage PHC credentials.
Their preferences depended on the type of ground truth data required for issuance, along with the familiarity, usability, and sensitivity of those data sources. They often considered facial recognition more resilient verification process than fingerprints.
 Other factors influencing preferences included the issuing ecosystem (centralized vs. decentralized), the issuer (government vs. private company), and the onboarding method (physical vs. remote).
Furthermore, we identified practical design suggestions to accommodate participants' needs, including periodic biometric checks and time-bound credential verification to ensure only the intended user accesses the credential; visually interactive human-checks to prevent social engineering during onboarding, collaboration between industry and government to establish decentralized standards for broader adoption.

\textbf{Contributions}
Overall, our study makes the following contributions: (1) Formative study sheds light on challenges in current personhood verification; (2) Our interview results provide rich insights into users' perceptions along with factors influencing their preference surrounding personhood verification; (3) User-Centered Design suggestions addressing from interactive sketch sessions.


%% file: Sections/2_Related_work.tex
\vspace{-2mm}
\section{Related Work}
\vspace{-2mm}
We outline the background of personhood credentials to introduce a foundational concept.
We then present the landscape of verification tools, including CAPTCHA, biometrics, and various other identification methods to discuss the limitation. Lastly, we discuss emerging practices of personhood verification to underscore the motivation for our work. 
\vspace{-2mm}
\subsection{Background on Personhood Credentials}
\vspace{-2mm}

\textbf{Proof of Personhood (PoP)} Verifying online identities has been particularly crucial for blockchain systems to mitigate the risks of impersonation and fraudulent activities. 
Proof of Personhood (PoP) has emerged to counter Sybil attacks, which manipulate peer-to-peer networks by operating multiple pseudonymous identities. It verifies an individual's humanness and uniqueness digitally on blockchain~\cite{borge2017proof} while preserving anonymity by linking virtual and physical identities.  
For instance, the Idena: proof of person blockchain runs a Turing test to prove the humanness and uniqueness of its participants~\cite{idenaWhitepaper}. 
Similarly, Humanode \cite{kavazi2021humanode, kavazi2023humanode}, which ensures one Human one Node, safeguarded by cryptographically bio-authorized nodes using 3D users’ faces. BrightID and Proof of Humanity leverage a social mechanism such as, a social graph and social vouching to achieve PoP \cite{shilina2023revolutionizing} and
guarantees that the users exist and are not duplicates of another entity. Recent identity verification platforms reflected such social needs for PoP; Gitcoin Passport, and Civic Passserves as a self-sovereign data collection protocol with a PoP algorithm~\cite{shilina2023revolutionizing}

\textbf{Personhood Credentials (PHCs)} Stemming from the concept of PoP, personhood credentials (PHCs) have been proposed as an alternative online identification credential against AI-powered deception \cite{adler2024personhood}. They define PHCs as \textit{``digital credentials that empower users to demonstrate that they are
real people—not AIs—to online services, without disclosing any personal information.''} The fundamental architecture of PHCs includes three entities: User; Issuer; Service providers. First, the user gets PHC from the issuer by providing evidence to verify their identity. Then, the user can use PHC across different service providers without providing identity evidence again. It enables users to demonstrate humanness to online service providers without disclosing any personal information. 
An example of PHC implemented in the application is Worldcoin, which is based on PoP and Zero-Knowledge Proofs (ZKPs) \cite{WorldWhitepaper, worldHumanness, de2024personhood}. It maintains privacy and anonymity via ZKPs while employing the human iris pattern as a conclusive biometric marker for PoP. 

\vspace{-2mm}
\subsection{Landscape of Verification Tools}
\vspace{-2mm}
\paragraph{CAPTCHA}
CAPTCHA \cite{von-Ahn2003-wr} has been the most common verification method to differentiate humans from bots. 
Early work 
emphasized CAPTCHA are often complicated for humans 
\cite{Bursztein2010-kf}. Similarly, Fidas et al. investigated the number of attempts needed to solve, finding that only 48.5\% solved CAPTCHA on the first try 
\cite{Fidas2011-ab}. 
Existing work provided a comprehensive review of methods and alternatives \cite{Moradi2015-is}, noting security concerns 
and accessibility issues. Designing CAPTCHA that is fully accessible to everyone is impossible since users have different levels of 
abilities. More recently, former literature 
highlighted participants had low satisfaction with invisible CAPTCHA despite less burden on users 
\cite{Tanthavech2019-dm}. 
From a security perspective, Kumar et al. reviewed various types of CAPTCHAs, showing that modern CAPTCHA-breaking techniques are highly successful, sometimes with a 100\% success rate and the majority of them over 50\% \cite{Kumar2022-uk}. 
Another study revealed bots outperform humans in terms of solving time and accuracy across different CAPTCHA types \cite{Searles2023-db}. All in all, the current CAPTCHA systems have limitations in achieving optimal usability and robust security.

\textbf{Biometrics.}
Biometrics is another typical method of verification. Recent studies highlighted its popularity among users due to its high usability and convenience, as represented by fingerprint and face unlock systems~\cite{mare2016study, De-Luca2015-mp, Bhagavatula2015-fi, Zimmermann2017-wr}. 
On the other hand, the authors shed light on usability issues in robustness to physical changes 
and external environments 
\cite{mare2016study, Bhagavatula2015-fi}. Moreover, prior research revealed that participants have more privacy concerns about biometric authentication methods\cite{Zimmermann2017-wr}. 
Wolf et al. found that non-experts tend to have lower security concerns about biometric authentication, which is possibly due to a lack of understanding \cite{Wolf2019-od}. A recent study demonstrated the cross-context predictability of biometric data, resulting in the security risk of biometric authentication \cite{Eberz2018-eg}. Their results showed that attackers can improve their impersonate performance by collecting biometric data from multiple sources. 

\textbf{Humanness verification.}
Selfies and video calls have become popular 
as liveliness tests to verify users' identity and exclude bots. 
Instagram introduced video selfies as a new option for users to verify their age, allowing age-appropriate experiences on the platform~\cite{instagramWaysVerify}. However, 
a Reddit post gained attention with the selfie and ID photo made with Stable Diffusion, which implies attackers can create deepfake ID selfies more easily than ever before \cite{techcrunchGenAICould}. Prior work \cite{hashmi2024unmasking} demonstrated human overall accuracy of detecting audiovisual deepfakes was only $65.64$\%, and people tend to overestimate their detection capabilities. Considering that deepfakes are becoming increasingly sophisticated, there are inherent limitations to using this method for verifying personhood.

\textbf{Economic identifiers.}
Various economic identifiers are widely used in our daily lives to connect multiple sources of information \cite{kennickell2016identity}. In the United States, 
Social Security Number and Employer Identification Number are used by the Internal Revenue Service to identify individuals and organizations for taxation and government operations. Implementing this identification number system ensured efficiency, fraud prevention, and fair tax obligations \cite{caplin1963taxpayer}. 
In other countries, identification systems have evolved in different ways reflecting social backgrounds- China's Social Credit System assesses their citizens based on social and economic activities for trust and compliance \cite{creemers2018china,cheung2022datafication}. In contrast, Estonia has developed a digital identity system, offering secure access to a wide range of services 
for economic growth and efficiency \cite{martens2010electronic,tammpuu2019transnational}.


\textbf{Digital identifiers.}
Among digital identifiers, public identifiers (e.g., phone numbers, email addresses) are the most popular methods in user verification. 
Two-factor authentication (2FA) is a way to ensure user authenticity by sending a one-time passcode (OTP) 
\cite{aloul2009two, sharma2024can}. However, 
obtaining multiple identifiers allows for the effortless creation of duplicate bot accounts. Devices also have personally identifying information, such as IP and MAC addresses, which can be used as a browser fingerprint, a unique identifier developed from the combination of client-side data \cite{laperdrix2020browser,zhang2022survey}. Recently, user-centric identity management has emerged as a key area of focus \cite{ahn2009privacy}. It respects users' autonomy as reflected in federated login like OpenID and OAuth \cite{recordon2006openid, hardt2012oauth, li2020user}. These developments introduced 
social login, where social media accounts can be used to log into third-party websites and apps \cite{gafni2014social}. Blockchain technology has been implemented into a user-centric identity system as self-sovereign identity (SSI) \cite{toth2019self}. 
SSI enables users to mutually authenticate via decentralized identifiers (DIDs) and verifiable credentials (VCs) to give individuals control over their digital identities without intermediaries \cite{muhle2018survey, brunner2020did}.

\textbf{Watermarking and Fingerprinting.}
Watermarking embeds invisible or visible markers into digital contents to identify whether it has been altered or forged \cite{van1994digital,mohanty1999digital,podilchuk2001digital}, whereas fingerprinting captures patterns 
from a set of collected data such as device characteristics and metadata \cite{qureshi2019blockchain}. These techniques are utilized in detecting deepfakes by verifying content authenticity and identifying 
manipulation or synthetic generation. Barrington et al. developed an “identity fingerprint”
to evaluate the authenticity of multimodal contents \cite{barrington2023single}. The former study also proposed a system to detect deepfake videos using a speech-based hybrid watermarking method \cite{qureshi2021detecting}. However, a recent study demonstrated the impossibility of strong watermarking for generative models 
and concluded verifying the source of the data is more critical 
in many disinformation cases \cite{zhang2023watermarks}. While digital artifact verification enhances content trustworthiness without affecting user experience, it often leads the discussion back to the need for direct user verification methods. 
\vspace{-2mm}
\subsection{Emerging Practices of Identity Verification}
\vspace{-2mm}
\label{subsec:verification_practice}
Online identity verifications are implemented across various sectors to meet context-dependent needs, including finance, healthcare, social media, government services, and emerging technologies like LLM applications.
Financial institutions implement advanced OCR and AI technologies to verify government-issued identification documents, particularly for the remote onboarding process for new customers, and utilize biometric systems like fingerprint and facial recognition \cite{yousefi2024digital}. In healthcare systems, health insurance cards have been widely adopted as an identification certificate with electronic versions also being developed \cite{chen2012non}. Biometrics-based verification also plays a significant role in this sector, 
as implemented robust fingerprint identification for electronic healthcare systems \cite{fatima2019biometric,jahan2017robust}. In this context, identity verification is crucial to protect patient privacy and ensure accurate delivery of medical treatments. 
As for social media platforms, 
Instagram tested video selfie and social vouching as new ways of age verification, in addition to uploading their IDs, to protect younger users and build safer online spaces \cite{instagramWaysVerify, instagramTypesID, metaTypesID}. 
In LLM applications, the current practice is simple verification through email addresses. However, there could be future cases where identity verification might be required to filter out malicious users.
Regarding government services, there is a verification service provided by the US federal government called Login.gov, which allows users to access online government services \cite{LogingovVerify, shovon2018restful}. It requires uploading a photo of an accepted driver’s license or state ID card. In recent years, many countries have introduced such verification solutions to streamline access to government services. 
For employment background checks, a tax identification card is sometimes regarded as a credential since it is often linked to employment history and income records, which aligns with this context. The fingerprint is also used to track criminal records \cite{cole2009suspect}. 

The evolving verification tools, in particular, emerging verification concepts like PHC highlight the need to address human factors often overlooked in technical research. Therefore, we aim to investigate these human aspects, providing insights into user perceptions and preferences that can be implemented into design implications.


%% file: Sections/3_Method.tex
\begin{figure*}[!t]
	\centering
	\includegraphics[width=\linewidth]{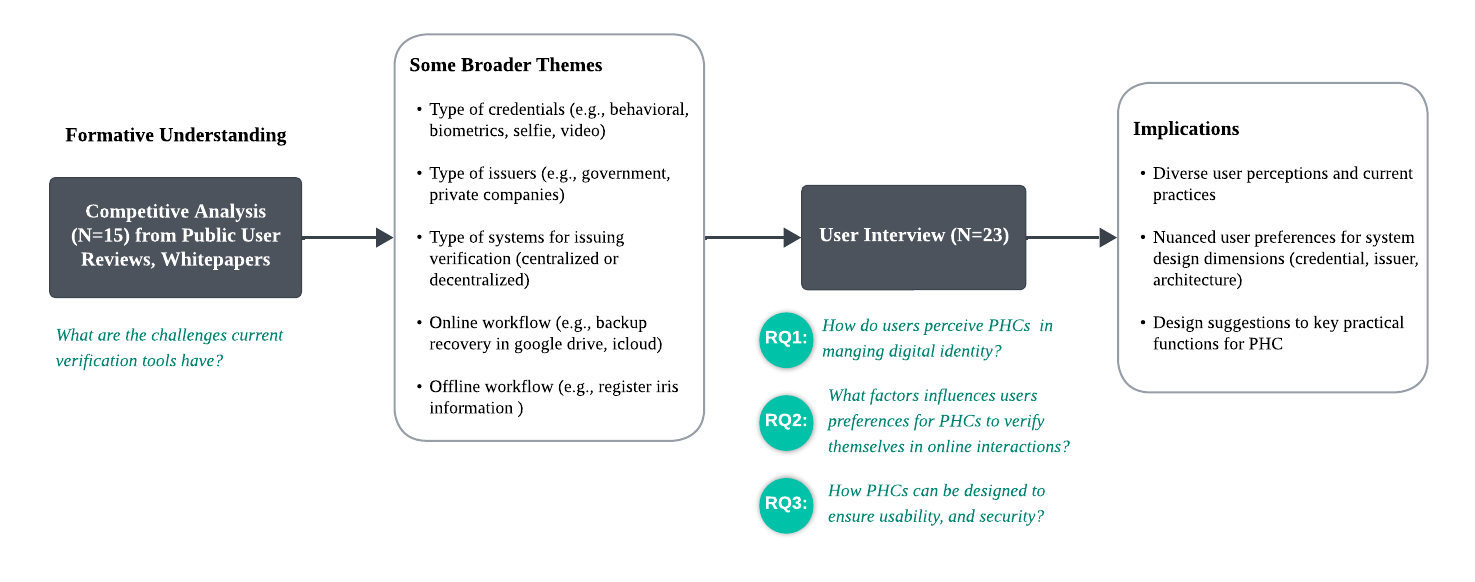}

	\caption{Method overview includes (a) a formative understanding of current personhood verification and related challenges through competitive analysis  (b) users' perception, preferences, and design through an interview study}
\label{fig:method}
\end{figure*}
\vspace{-2mm}
\section{Method Overview}
\label{sec:method}
\vspace{-2mm}
Building on the existing literature, it is clear that while significant progress has been made, a critical gap remains in understanding the key factors to operationalize personhood credentials that balance privacy, security, and trustworthiness online. 
As outlined in Figure~\ref{fig:method}, our study comprises: (1) a competitive analysis of current personhood/identity verification tools to identify challenges. These insights inform the design of a user study aimed at (2) investigating users’ perceptions (RQ1), identifying factors influencing their preferences for personhood credentials (RQ2), and conceptualizing designs (RQ3) to address these challenges.


\vspace{-2mm}
\section{Formative Understanding of PHCs}
\vspace{-2mm}
In this section, we outline our formative analysis of existing personhood verification systems, which informed the design rationale for developing our user study (Section~\ref{user-study}).

We systematically consolidated a list of systems based on their popularity, diversity in platform type (centralized vs. decentralized), and relevance to the domain of digital identity~\cite{idenaWhitepaper, kavazi2021humanode, kavazi2023humanode, de2024personhood, BrightID, PoH, adler2024personhood}
This consists of
World app, BrightID, Proof of Humanity, Gitcoin Passport, and Federated Identities (OAuth), etc (Table~\ref{tab:systems}). 
Table~\ref{tab:identity_verification} provides an overview of different attributes of how existing systems operate and their design trade-offs. We found 15 apps categorized into six groups. Five of these were centralized, primarily government-based personhood verification systems. This initial categorization is based on the data requirements for issuing credentials varied, including behavior filters, biometrics (such as face, selfie, iris, or video), social graph and vouching mechanisms, physical ID verification, and, in some cases, combinations of these methods. 
We also documented on how users navigate the system and identify potential usability and security issues. Two UI/UX in out team evaluated whether users could successfully sign up and obtain personhood credentials. We independently compiled an initial list of evaluation results based on key questions. This includes- \textit{``How intuitive is the verification process?; How effectively does the platform provide feedback during different steps of registration and verification?; How do we as users feel regarding the data requirements in the verification systems?; How does the platform manage users' data?; What are the potential risks regarding users' privacy in the platform?''}
Given the limited access to systems like Estonia’s digital ID, Civic, and China’s social credit system, we used available white papers and documentation to reconstruct their workflows. Finally, we synthesized our observations and conducted qualitative coding to identify recurring themes.

\begin{table}[ht]
    \centering
    \scriptsize
    \begin{tabular}{llll}
      \hline
       App Name  & Source & reviews  \\
    
        \hline
     Worldapp & Documentation~\cite{WorldWhitepaper}, Google Play Store& 1523 \\
  BrightID & Documentation~\cite{BrightID},Google Play Store & 328 \\
  DECO & Documentation~\cite{zhang2020deco} & Review  \\
  CANDID & Documentation~\cite{maram2021candid} & Review \\
  Proof of Humanity &  Documentation~\cite{PoHexplainer} & Review \\
  Adhar Card &  Documentation~\cite{Aadhaar}, Google Play Store & Review
  \\
Estonia e-ID  &  Documentation~\cite{estoniaE-ID} & Review\\
Chinese Credit system &  Documentation~\cite{ChinaSocialCreditSystem} & Review \\
Japan My Number Card &  Documentation~\cite{JapanMyIDNumber} & Review \\
ID.me &  Documentation~\cite{irsIdentityVerification, idAccessAll}, Google Play Store & Review \\
Idena &  Documentation~\cite{idenaWhitepaper} &  Review \\
Humanode &  Documentation~\cite{kavazi2021humanode} &Review\\
Civic &  Documentation~\cite{CivicPass} &Review \\
Federated identities (Oauth) &  Documentation~\cite{OAuth} & Review\\
  \hline
    
    \end{tabular}
    \caption{Competitive Analysis Data Sources 
    }
    \label{tab:systems}
\end{table}

\begin{table*}[h!]
    \centering
    \caption{Comparison of Existing Personhood Verification Systems}
    \label{tab:identity_verification}
    \resizebox{\textwidth}{!}{ 
    \begin{tabular}{l >{\small}l >{\small}l >{\small}l >{\small}p{3cm} >{\small}p{2.5cm} >{\small}l} 
        \hline
        \textbf{Category} & \textbf{Service Name} & \textbf{Architecture} & \textbf{Issuer} & \textbf{Credential} & \textbf{Platform} & \textbf{Free/Paid} \\
        \hline
        \hline
        \multirow{3}{*}{Behavioral Filter} 
        & CAPTCHA & Centralized & open-source, vendor & Recognize distorted texts, images, sounds etc. & Desktop and mobile browsers & Free/Paid\\
        & reCAPTCHA & Centralized & Google & Click checkbox & Desktop and mobile browsers& Free/Paid\\
        & Idena & Decentralized & open-source & Solve contextual puzzle & Blockchain & Free\\
        \hline
        \multirow{2}{*}{Biometrics}
        & World ID & Decentralized & World & Biometrics (iris scan) & App (iOS, Android) & Free\\
        & Humanode & Decentralized & Humanode & Biometrics (face) & Blockchain & Paid\\
        \hline
        Social Graph 
        & BrightID & Decentralized & open-source & Analysis of social graph & App (iOS, Android) & Free\\
        \hline
        Social Vouching 
        & Proof of Humanity & Decentralized & Kleros & Social vouching & Web & Paid\\
        \hline
        \multirow{2}{*}{Decentralized Oracle} 
        & DECO & Decentralized & Chainlink Labs & Cryptographic proof & Decentralized oracle & Under PoC\\
        & CANDID & Decentralized & IC3 research team & Cryptographic proof & Decentralized oracle & Under PoC\\
        \hline
        \multirow{4}{*}{Government-based ID} 
        & India Aadhaar Card & Centralized & Government & Document-based or Head Of Family-based enrollment + digital photo of face, 2 iris, and 10 fingerprints& Web, App (iOS, Android) & Free\\
        & Estonia e-ID & Decentralized & Government & Passport or EU ID + digital photo of face & Web, App (iOS, Android) & Paid\\
        & Japan My Number Card & Centralized & Government & Issue notice letter + photo ID or two non-photo IDs & Web, App (iOS, Android) & Free\\
        \hline
        \multirow{2}{*}{Others} 
        & ID.me & Centralized & ID.me & Government-issued ID & Web & Free\\
        & Civic Pass & Decentralized & Civic & Government-issued ID, Biometrics (face), Humanness, Liveness & Web & Free\\
        \hline
    \end{tabular}
    }
\end{table*}

\begin{figure*}[h]
    \centering
    \begin{subfigure}{0.48\textwidth}
        \centering
        \raisebox{0.5\height}{
        \includegraphics[width=\textwidth]{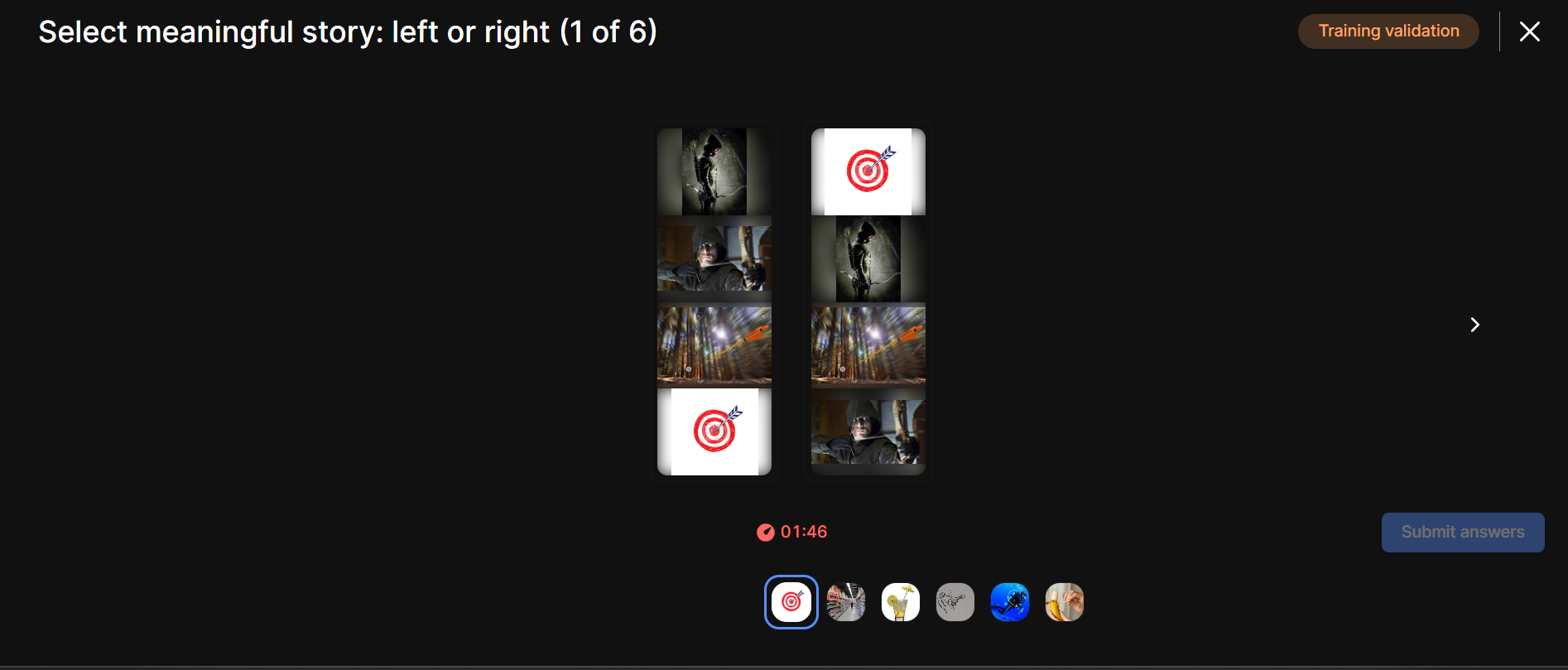}}
        \captionsetup{width=\textwidth, font=footnotesize} 
        \caption{Idena validation test interface: This requires users to select meaningful stories within a time limit, which can pose challenges for new users}
        \label{fig:idena}
    \end{subfigure}
    \hfill
    \begin{subfigure}{0.48\textwidth}
        \centering
        \includegraphics[width=\textwidth]{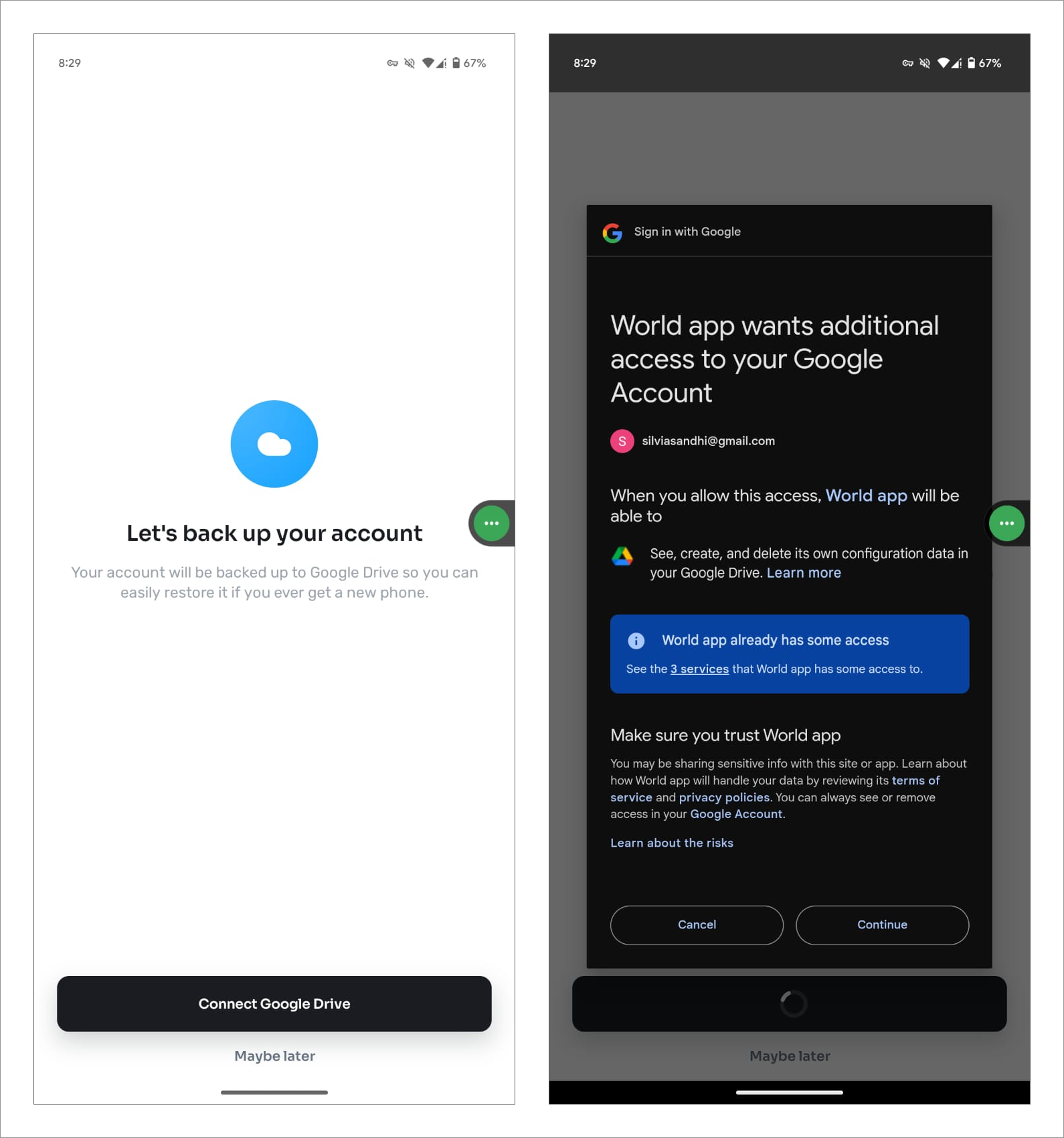}
        \captionsetup{width=\textwidth, font=footnotesize} 
        \caption{World App backup interface: requires users to connect Google Drive}
        \label{fig:worldapp}
    \end{subfigure}
    
    \caption{PHC-related interfaces: (a) Idena validation test, (b) World App backup process.}
    \label{fig:phc_interfaces}
\end{figure*}

\vspace{-2mm}
\subsection{Challenges in Identity Verification}
\vspace{-2mm}
\label{prac-cha}
\textbf{Demanding Cognitive and Social Efforts for Verification Workflow.}
We found platforms such as World App and BrightID developed on decentralized technologies, 
including zero-knowledge proofs and social connections, may confuse non-technical users. For instance, user review from playstore suggested-many having issues understanding how to receive BrightID scores to prove they are sufficiently connected with others and verified within the graph. In their words \textit{``It's hard for me to connect with people to create the social graph.''} 
From experts' evaluation of UI/UX, we found Proof of Humanity lacks options to correct or update mistakes, which can make the registration process less user-friendly. 
Similarly, Idena's validation test (flip test) (Figure~\ref{fig:idena}) was challenging as new users as it required to create a meaningful story within the allotted time and earn enough points for validation. Simialrly, World App's(Figure~\ref{fig:worldapp}) account creation process to get an identifier doesn't inform users how and why to navigate the app can undermine intended functionality,  or underutilization of the app’s capabilities.


\textbf{New or Complex System Rule to Recover ID. }
Both from UI/UX task and playstore review, we found the BrighID recovery process tedious and the rules unclear. A representative user review stated-\textit{``If you create an account and do not set up recovery connections you cannot get your account back. This forces you to create a new account which defeats the purpose of the app.''}
Another workflow of World App that requires users to connect their Google Drive to back up their accounts. However, this process may confuse users and create challenges during account recovery if they fail to complete the backup(Figure~\ref{fig:worldapp}).


\textbf{Privacy and Data Requirement Issue. }
From our competitive analysis (Table~\ref{tab:litcomparison}), Data requirements across the systems vary significantly in scope and sensitivity. Decentralized platforms like World App, and BrightID required minimal data collection to issue ID while Proof of Humanity require video submission to receive a credential for was quite invasive when the videos were open to the public with clear faces.
While there is benefit of decentralization, often it is not clear how exactly service providers will handle the data in their policies and white papers.
In contrast, Federated Identities OAuth\cite{OAuth} login process streamlines and this contributed to using known third-party service providers. This ensures ease of use as users need to specify the identity provider during the login or authentication process and grant access to their specific data. This reflects the importance of known entities and level of trust in data handling.
Centralized systems, including Aadhaar and Estonia digital ID, require extensive personal and biometric data—fingerprints and iris scans—to ensure verification services while experts expressed privacy concerns towards china’s Social Credit personhood System, especially the use of it in measuring social scores.

\textbf{Requirement of Optimal Device or Physical Presence.}\\
Government-supported systems like Aadhaar and Estonia e-Card feature structured interfaces but come with limitations: Aadhaar’s biometric registration may challenge rural populations, while Estonia’s dependence on smart-card hardware might exclude those without the necessary devices. Proof of Humanity, Humanode, Civic Pass may create challenges as proper lighting and optimal devices are necessary for taking the appropriate photo or video for biometric verification
On the contrast, Aadhaar card\cite{Aadhaar}\cite{AadhaarEnrollment}, Estonia's e-ID and Japan's My Number Card require one to be physically present and the issuing process takes a long time can create user frustration. 
\vspace{-2mm}
\section{ User Study Method}
\vspace{-2mm}
\label{user-study}
This section outlines the method for exploring users' perceptions and preferences of personhood credentials. We conducted semi-structured interviews with 23 participants from the US, and the EU/UK in October 2024.
The study was approved by the Institutional Review Board (IRB).
\vspace{-2mm}
\subsection{Participant Recruitment}
\vspace{-2mm}
We recruited participants through (1) social media posts, (2) online crowdsourcing platforms, including CloudResearch and Prolific. Respondents were invited to our study if they met the selection criteria: a) 18 years or older and b) living in the US or the EU/UK. Participation was voluntary, and participants were allowed to quit anytime. Each participant received a \$30 Amazon e-gift card upon completing an hour-and-a-half interview.

\subsection{Participants}
We interviewed 23 participants, 10 from the US and 12 from the EU/UK. The majority of the participants (61\%) were in the age range of 25-34, followed by 22\% were 35-44 years old. The participants were from the United States and various countries, namely Spain, Sweden, Germany, Hungary, and the United Kingdom. Participants had different backgrounds of education levels, with 87\% of participants holding a Bachelor’s degree and 65\% holding a graduate degree. 65\% of participants had a technology background, while 48\% of them had a CS background. All participants reported using online services that required them to verify their personhood. Table~\ref{table:demographics} presents the demographics of our participants. We refer to participants as P1,. . . ,23.
\begin{table*}[h!]
\centering
\caption{Overview of PHC Application Scenarios}
\label{table:scenario}
\begin{tabular}{lll}
\hline
\textbf{Scenario} & \textbf{Service} & \textbf{Credential} \\
\hline
Financial service & Bank, Financial institutions & Passport or Driver’s license, Face scan \cite{yousefi2024digital}\\
Healthcare service & Hospitals, Clinics & Health insurance card,  Fingerprint \cite{chen2012non,fatima2019biometric,jahan2017robust}\\
Social media & Tech companies & National identity card, Video selfie \cite{instagramWaysVerify, metaTypesID,instagramTypesID} \\
LLM application & Tech companies & Iris scan \cite{WorldWhitepaper, worldHumanness}\\
Government service & Government & Driver’s license or National identity card \cite{LogingovVerify}\\
Employment background check & Background check companies & Tax identification card, Fingerprint\cite{cole2009suspect}\\
\hline
\end{tabular}%
\label{tab:scenarios}
\end{table*}
\begin{table*}[h]
\centering
\caption{Participant demographics and background.}
\resizebox{\textwidth}{!}{%
\begin{tabular}{l l l l l l l l}
\hline
\textit{Participant ID} & \textit{Gender} & \textit{Age} & \textit{Country of residence} & \textit{Education} & \textit{Technology background}  & \textit{CS background} &\textit{Residency duration} \\
\hline
P1 & Male & 25-34 & the US & Master's degree & Yes & Yes &3-5 years\\
P2 & Female & 25-34 & the US & Master's degree & Yes & Yes & 1-3 years\\
P3 & Female & 25-34 & the UK & Master's degree & Yes & No & 1-3 years\\
P4 & Female & 35-44 & the UK & Some college, but no degree & Yes & Yes & Over 10 years \\
P5 & Male & 25-34 & the US & Doctoral degree & Yes & Yes & 5-10 years \\
P6 & Male & 35-44 & the US & Less than a high school diploma & No & No & Over 10 years \\
P7 & Male & 25-34 & the US & Doctoral degree & Yes & Yes & 3-5 years\\
P8 & Male & 45-54 & the US & Bachelor's degree & Yes & Yes & Over 10 years \\
P9 & Female & 25-34 & New Zealand & Master's degree & No  &  No &  Over 10 years\\
P10 & Male & 25-34 & the US & Master's degree & No & No & Over 10 years\\
P11 & Female & 25-34 & the UK & Bachelor's degree & No & No & Over 10 years\\
P12 & Male & 18-24 & the UK & Master's degree & Yes & Yes & 1-3 years\\
P13 & Male & 35-44 & the UK & Bachelor's degree & Yes & No & Over 10 years\\
P14 & Male & 25-34 & Sweden & High school graduate & No & No & Over 10 years \\
P15 & Female & 25-34 & Spain & Master's degree & Yes & Yes & Over 10 years \\
P16 & Female & 25-34 & Germany & Master's degree & Yes & Yes & Over 10 years \\
P17 & Female & 25-34 & Spain & Doctoral degree & No & No & Over 10 years \\
P18 & Female & 35-44 & the US & Bachelor's degree & No & No & Over 10 years \\
P19 & Female & 25-34 & Germany & Master's degree & Yes & Yes & 3-5 years \\
P20 & Male & 25-34 & Hungary & Master's degree & Yes & No & 3-5 years \\
P21 & Male & 35-44 & the US & Bachelor's degree & Yes & No & 5-10 years \\
P22 & Female & 18-24 & France & Master's degree & Yes & Yes & Less than 1 year\\
P23 & Male & 45-52 & the US & Master's degree & No & No & Over 10 years\\
\hline
\end{tabular}%
}
\label{table:demographics}
\end{table*}

\vspace{-2mm}
\subsection{Semi-Structured Interview Procedure} \label{sec:study_protocol}
\vspace{-2mm}

We started with a round of pilot 
studies (n=5) to validate the interview protocol. Based on the findings of pilot studies, we revised the interview protocol.

\textbf{Open Ended Discussion.} We designed the interview script based on our research questions outlined in the introduction section~\ref{sec:introduction}. 
At the beginning of the study, we received the participants’ consent to conduct the study. Once they agreed, we proceeded with a semi-structured interview. The study protocol was structured according to the following sections: (1) Current practices regarding digital identity verification; (2) Users' perception of PHC before and after watching the informational video; (3) Scenario-based session to investigate factors that influence users' preferences of PHC; 
(4) Design session to conceptualize users' expectations; (5) A brief post-survey on Users' Preference of PHCs.

In the first section, we first asked a set of questions to understand participants' current practices of online platforms and the types of identity verification methods they had experience with. This is to understand their familiarity with different types of verification, such as biometrics, physical IDs, etc.

In the second section, we then asked about participants' current understanding and perception of personhood credentials either from prior knowledge or from intuition by just hearing the term. 
While the majority recognized this as unfamiliar terminology, most inferred that it referred to a form of personal identification, often associating it with biometric verification.
Then, we showed them an introduction video on PHC \footnote{https://anonymous.4open.science/r/PHC-user-study-14BB/}, 
covering their definition, 
and implications of it in online services. Based on former literature\cite{adler2024personhood}, we designed the video with easy-to-understand text, visuals, and audio to make the concepts accessible to average users. We created a set of knowledge questions to assess participants' understanding of PHC before and after showing the video. 

For instance, we observed an improve in correct response rate for the question, such as, \textit{``What is the primary goal of PHC?''} from 85\% to 100\% after watching the video.
In the third section, we focused on scenario-based discussions, exploring specific applications of PHC to understand factors that influence participants' preferences towards PHCs as well as identify challenges to leverage in PHC design for various services. We examined the following six scenarios: (1) Financial service, (2) Healthcare service, (3) Social Media, (4) LLM applications, (5) Government Portal, and (6) Employment Background Check.
We have also incorporated various types of data or credentials requirements (e.g. physical id, biometrics, etc) across scenarios to maintain diversity in our discussion with participants as shown in Table.\ref{table:scenario}. 
We selected types of credentials for each scenario based on former literature and existing PHC as explained in the section \ref{subsec:verification_practice}. 

For each of the six scenarios, we explored participants' perceptions of using PHC in hypothetical situations that align with the research focus as well as to help participants can relate PHC concepts to real-world applications. This is particularly useful for this study where where user perceptions and expectation under specific conditions are crucial to devising solutions \cite{carroll2003making}.
We asked about their feelings, perceived benefit and risks. We also nudge them to think about any privacy and security perception around using PHC and types of data (e.g., iris, face, government id, etc) involved in issuing PHC. 

\fixme{
}

In the fourth section, we began by refreshing participants’ memories of the various risks and concerns discussed in the earlier scenario-based section. Following this, we guided participants to brainstorm potential design solutions by sketching their ideas to address these concerns. To facilitate the sketching process, we developed sketch notes in Zoom as prompts to help participants generate ideas, particularly when starting from scratch is challenging. 
We also investigated participants' preferences for PHC regarding the issuers and issuance systems of PHCs, as well as the types of data required for issuing PHCs. 
We encourage participants to explain their reasoning. These questions were informed by insights from the pilot study, where participants expressed preferences for different types of data, system architecture, and various stakeholders involved in PHC issuing.

\textbf{Post-Survey.}
We conducted a post-survey to obtain participants' PHC preference quantitatively. It included questions on participants' preference on credential type, issuer and issuance system  for the scenarios (e.g., financial, medical, etc) we considered in our interview.

\vspace{-2mm}
\subsection{Data Analysis}
\vspace{-2mm}
Once we got permission from the participants, we obtained interview data through the audio recording and transcription on Zoom. We analyzed these transcribed scripts through thematic analysis \cite{Braun2012-sz, Fereday2006-yv}. Firstly, all of the pilot interview data was coded by two researchers independently. Then, we compared and developed new codes until we got a consistent codebook. Following this, both coders coded 20\% of the interview data of the main study. We finalized the codebook by discussing the coding to reach agreements. Lastly, we divided the remaining data and coded them. After both researchers completed coding for all interviews, they cross-checked each other’s coded transcripts and found no inconsistencies. Lower-level codes were then grouped into sub-themes, from which main themes were identified. Lastly, these codes were organized into broader categories. Our inter-coder reliability (0.90) indicated a reasonable agreement between the researchers.

%% file: Sections/Results_RQ1.tex
\vspace{-2mm}
\section{RQ1: Users’ Current Impression of PHC}
\vspace{-2mm}

\textbf{Current Verification Practices.} 
When discussing identity[personhood] verification, 
participants most commonly mentioned financial services, including online banking and investment platforms, as well as health services, government-related processes, and cross-border regional verifications in both
digital and physical forms of credentials. Several participants mentioned they are required to upload government IDs (e.g., social security number, driver's license) when creating an account for financial services- as P7 said--\textit{``For Robinhood, it asked for uploading my government-issued IDs like driver's license and passport.''} 
Similarly, P2 explained their verification experience in the government services requiring multiple information \textit{``Last year when I was requesting my tax filing documents in IRS. To access them, I had to verify my identity with my face, as well as information from government-issued IDs to confirm my identity. [id.me~\cite{irsIdentityVerification}].''} Beyond financial and government services, identity verification has also become essential in marketplace apps,
P1 illustrated this trend with Airbnb, explaining  \textit{``If you do book an Airbnb, at least as a renter, you would need to verify your government id before being eligible to book your first stay.''} 

\textbf{Trade-offs between known and unknown privacy guarantee.} 
Participants often made trade-offs between familiar security guarantees associated with traditional verification methods over the less clear assurances of emerging PHC. As P18 mentioned she heard about World ID as a personhood credential and remarked \textit{``
They scan your iris, create a unique code, and re-verify me again later time. I guess that means they still keep some sort of data from the iris scan. I would still stay with email or traditional verification as I know what they are keeping. Even though world app guarantees privacy with hi-tech, I don’t exactly know how much privacy I have in human tests. I’m not saying it’s a bad idea, it's just i am unsure.''} 
P23 with age ranges from 45-52 noted--\textit{``it's just easier to do with email or my physical photo id in an old system 
Newer technology is supposed to be faster and more user-friendly, but to me, iris scan or selfie scan, I can't even know if I am doing it correctly. I’m getting to the point where I can say I’m old-fashioned.''} This underscores the generational divide in preferences with emerging technologies compared to established methods.

\textbf{Perceived Benefits of PHC: Fairness in Representation.} Participants discussed fairness and the potential benefits of PHCs. P19 shared her experience with online study platforms which is one of the primary means of earning, \textit{``often time I got "time out or returned" in this platform and earlier I thought I am slow responding to survey request I received on my account. Lately, it happens too often, feels like there are bots or a group of people are more proactive in participating in surveys. honestly, it reduces the chances for me being selected for research studies ''} She emphasized that personhood verification could help filter out bots and increase her chances as a genuine participant.  P21 who is a dedicated gamer who find it disheartening to encounter players, likely bots, who level up so quickly. To emphasize the benefit of PHC, he mentioned \textit{``“ I play online games with other people around the world, like shooting games. We regularly encounter bots in these games, which basically dilutes our experience. As an experienced player, I can usually tell when someone isn’t a real person
based on how they play. personhood verification could be a good feature.''}


\textbf{Perceived Concerns: Power, Control, Security} 
Participants expressed concerns about potential risks, including the centralized power of the issuer, uncertain regulations of emerging technology, authenticity of PHC, and misuse of anonymity.
For instance, P6 pointed out that centralization of credential information with PHC issuers could lead to power being misused, saying, \textit{"you're gonna give all your information to a small group or institutions to issue PHC. So they have the power that can be abused later."} 
P10 also highlighted the risks in high-stakes situations, such as employment background verification, where an inauthentic PHC could have serious consequences. He noted \textit{"I don't know if it's possible to fake government-issued ID. I feel like I'm probably concerned about an inauthentic PHC."} P4 in a similar topic expressed concerns about how much data will be revealed to an employer if they only share PHC \textit{``For instance, I had a job offer that required details about my criminal record from five years ago. It would probably reveal more even if I only share PHC. how much anonymity I have.''}

\textbf{PHC Preference Dilemma: Physical and Digital Verification.}
Many participants often relate their offline experiences when describing their preference for credentials and issuance systems.
For instance, some of them described the inconvenience of carrying physical IDs 
, saying \textit{``When you need to buy some alcohol drinks, take a flight, you need to show your driver's license or passport. 
But it's not a very convenient, because I have to carry the physical ID all the time.''} Another participant pointed out the potential risk of IDs being stolen, citing, \textit{``I think this information may be lost, or maybe stolen. So it's a risk.''} 
While participants highlighted the inconvenience and physical vulnerabilities of traditional IDs, they simultaneously shared the limitations of digital identifiers, such as verification through phone number [OTP] during international travels.

%% file: Sections/Results_RQ2.tex
\vspace{-2mm}
\section{RQ2: Factors Influence on How People Would like to Verify Themselves }
\vspace{-2mm}
In this section, we discuss various factors that influence people's preferences including application type, credential types, stakeholders as issuers, and architecture type. 
\vspace{-2mm}
\subsection{PHC Onboarding: Online Vs Offline}
\vspace{-2mm}
Participants identified the onboarding process as a key factor in managing PHCs. While some preferred a hybrid approach over a fully online system, others favored a fully online process for its convenience. Many drew parallels to how they opened their first bank accounts, emphasizing the importance of in-person verification. They described visiting the bank, presenting their IDs and passports to a bank official, and then receiving their account. Reflecting on that experience, participants indicated that PHCs, especially for financial services, should be issued in a similar way. One of the perceptions was the high risk of security of a fully online system where user would use their own browser at home to upload certain ids and passports to receive PHC compared to bank officials doing the same thing in their protected system. P22 said \textit{``There is a high risk with a fully online system where users would use their own browsers at home to upload certain IDs and passports to receive a PHC. Compared to that, having a bank official verify the information in their protected system feels much safer.''} 
However, some highlighted the impracticalities of in-person verification as P18 mentioned \textit{``I live in west coast, now i need to travel to a location to have my irises scanned by the Orb device, which is not practical for me.''}

\vspace{-2mm}
\subsection{Data Requirements to Issue Credential}
\begin{figure}[!t]
    \centering
    \begin{subfigure}{0.48\linewidth}
        \centering
        \includegraphics[width=\linewidth]{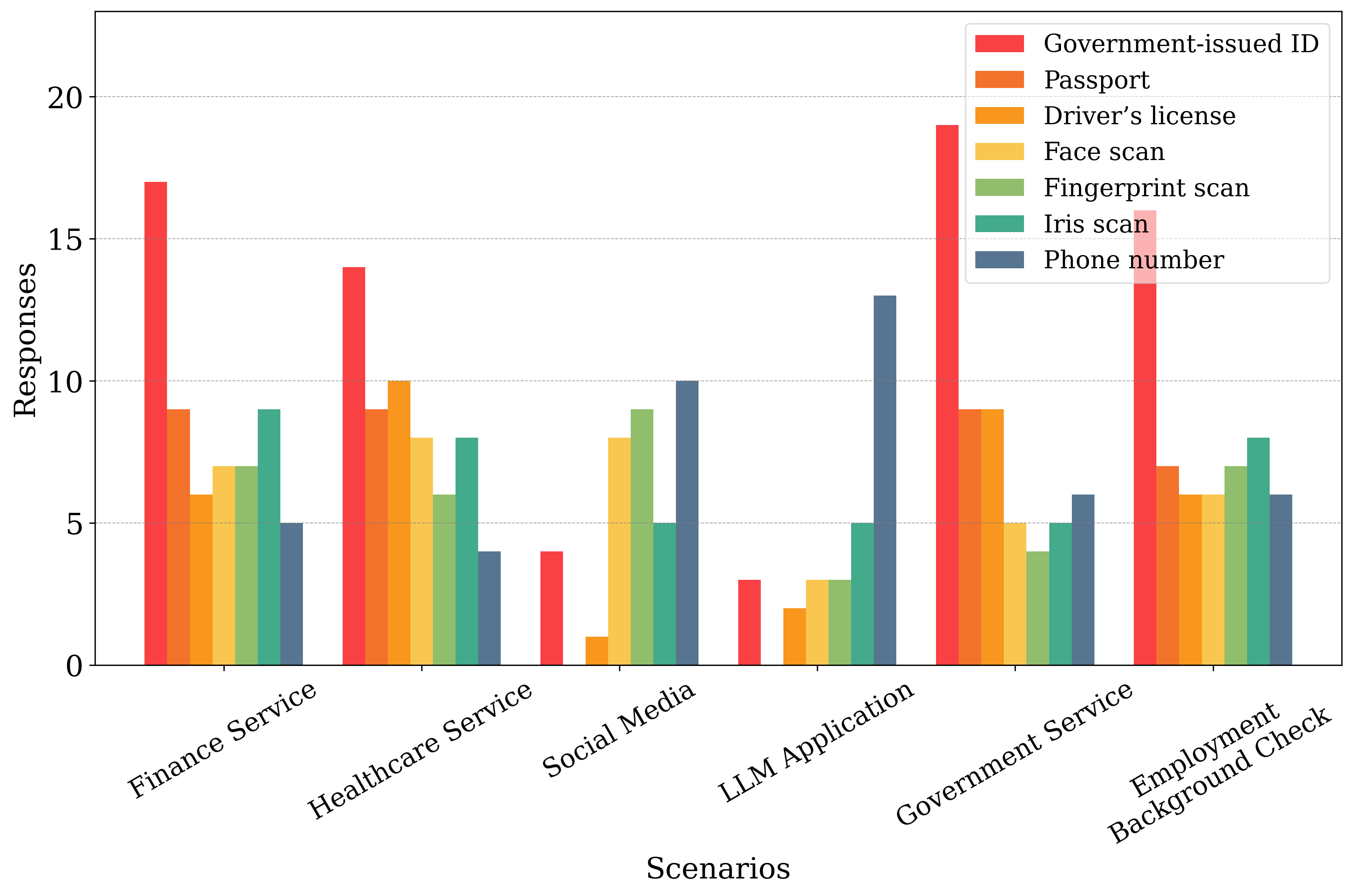}
        \caption{Results of credential preference: Which types of credential would you prefer to use as personhood verification for each scenario? (Multiple selections were allowed.) }
        \label{fig:credential}
    \end{subfigure}
    \hfill
    \begin{subfigure}{0.48\linewidth}
        \centering
        \includegraphics[width=\linewidth]{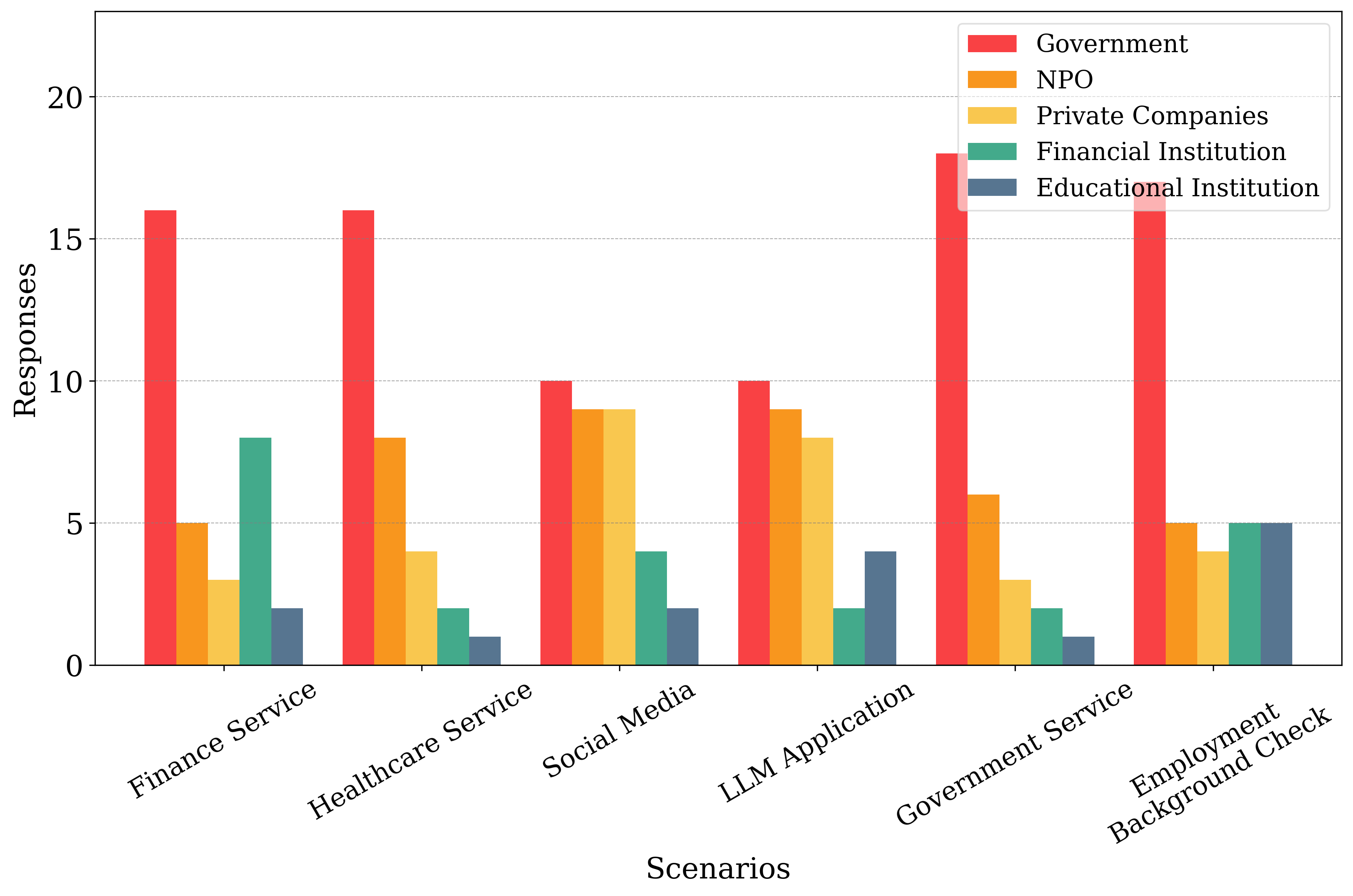}
        \caption{Results of issuer preference: Which type of issuer would you prefer to issue and manage your PHC for each scenario (Multiple selections were allowed.)}
        \label{fig:issuer}
    \end{subfigure}
    \vspace{-10pt}
\end{figure}

\vspace{-2mm}
Data requirement in issuing PHC credentials was one of the main factors. Participants generally categorized data requirements into three main types: \ding{202} Government-issued IDs, such as social security numbers, passports, and driver's licenses; \ding{203} Biometrics, including face, fingerprints, and iris; and \ding{204} Digital identifiers, such as phone numbers and email addresses. To grasp their preferences comprehensively, Figure \ref{fig:credential} presents quantitative results of credential preference from post-survey.
For instance, a government-issued ID is the most preferred data to issue PHC across applications except social media and llm applications. Phone number is most preferred in llm application followed by Iris scan.


\textbf{Familiarity with traditional (e.g. govt id)} 
Several participants highlighted their preference for their PHC associating with government-issued IDs as they were most familiar with this approach and perceived it as reliable.
To add some complexity, P11 mentioned the practicability of different types of government id- \textit{``Driver's license is pretty common, and  we usually bring that all the time. It's easy to bring and easy to take a photo and upload. A passport or a social security number is not the one that people usually bring. So if we need to verify that we 1st need to come back home and then search it, search them and then provide it. So it takes, additional steps. But a driver's license pretty easy.''} 

\textbf{Sensitivity, Security, \& Efficiency Across Different Biometrics}
We found that participants favored biometrics due to their ease of use, no requirement to remember passwords minimized exposure of personal data, and their functionality as standalone credentials. Highlighting efficacy and privacy P4 stated - \textit{``government id is gonna have a lot of your personal information. biometrics here minimize data-just biomarker. like with iris or fingerprint, it's just the minimal and still being secure.''} 
We also found that individuals have varying preferences across biometrics types, such as fingerprints, facial recognition, selfie, and iris scanning. Firstly, participants considered fingerprints as less invasive and less sensitive.
In contrast, facial recognition was discussed as a more sensitive biometric method, as illustrated by P5, \textit{"I would say fingerprint is fine, but face is too much. you can be identified in public settings in streets. If they have camera [surveillance], anyone can just trace your whereabouts."}  Regarding iris recognition, most participants expressed their views without having direct experience with this method. Notably, P8 said-
\textit{"iris verification is probably the more secure of all of the biometrics. Because I know that fingerprints can be regenerated, and facial recognition can be regenerated. But irises are not easily generated. The eye is a very complex organ in the body. So if I had a choice between the 3 forms of biometrics, I would choose iris, because I think it's the most secure."} In content of efficacy, P7 emphasized the efficiency of face compared to fingerprints, stating \textit{"I think facial recognition is better and more technologically improved, it is quick.''} On the same note, 
P23 highlighted challenges with their father fingerprint, noting that it had become difficult to identify due to the type of construction work they engaged in, causing the prints to blur over time.

\textbf{Combination of Multiple Credentials}
Participants emphasized the importance of using multiple type of credentials to enhance security. P13 indicated this preference, \textit{combination of facial scan and fingerprint combination. If you're wanting to prove this is the person. Then it would make sense to have more than one biometric. I don't know if it's possible to fake somebody's. If there are completely two different biometrics, it would be more difficult to fake.''} In cases like financial service, many participants preferred PHC issuance based on both physical id (e.g. gov id) and biometric (e.g. iris).

\vspace{-2mm}
\subsection{Stakeholder Types}
\vspace{-2mm}

\textbf{Control and Practicality: Preferences of PHC Issuers.}
Participants expressed varied levels of acceptance regarding the preferred issuers of PHC issuers. Across all application scenarios (Figure~\ref{fig:issuer}), government entities emerged as the most trusted issuers for the majority of participants, followed by nonprofit organizations (NPOs). P5 noted-- \textit{"Government is my preferred. IIf there are certain organizations are leveraging 3rd party organizations, they should be regulated and under government supervision. The least favored one is private companies without supervision, like commercial companies doing their own."} 
However, practical considerations influenced participants' views in certain domains, particularly social media and LLM applications, where private companies were rated as acceptable issuers by some. As P19 said \textit{``I don't see it happening where social media will involve govt vetted PHC. its just not practical. and truthfully, i don't want gov issued phc for social media, i am not fool to allow govt to another layer of surveillance.''} In contrast, a group of participants favored nonprofit organizations for domains like healthcare and social media, valuing their balance of trust and regulation --P10 stating \textit{``So NPOs I feel like they do have that government backing, they could be another trustworthy source, but not as intense as the govt. to balance it out.''}This contrast underscores the complex trade-offs individuals consider when evaluating trust, regulatory oversight, and practicality in selecting entities to issue PHCs. 
\textbf{Trustworthiness of Stakeholders}
When discussing the current verification process, participants often indicated that there are organizations they can trust and others they cannot. 
For instance,  Some worried whether these issuers could be trusted to protect their personal information and maintain their privacy. P7 described these concerns and discussed how trust in issuers can be a deciding factor. \textit{"My only concern is, how do we trust the PHC am I using it? Because they are the entity who handles all these information, and they are collecting everything about us from the government...I don't want to share my personal information anymore, with some random website...But how the way the issuer can be understood? I think that would be the decision factor for users to start engaging with the system."}
Trustworthiness are associated with various aspects of stakeholders like regulations, customer support, past experiences with issues.
P5 explained his trust in banks comes from regulatory aspects: \textit{"They are being monitored by federal agencies. Their activities are monitored. They are under a lot of regulations. So there is a monitoring system that is tracking banks."} P4 emphasized her trust in banks with accessible customer support, which allows individuals to review the necessary steps and address any concerns they might have.
On the other hand, we observed that many participants expressed distrust towards social media platforms. P3 shared an experience of her social media account almost being hacked, 
noting \textit{"Someone tried to hack into my account...I've had people send me really dodgy links, which I knew instantly, that if I clicked on that link, it would mean that my account would get hacked as well."} 
\vspace{-2mm}
\subsection{Architecture Types}
\vspace{-2mm}
We observed diverse preferences for PHC architectures, with participants highlighting decentralized models for improving centralized data security and enabling user choice of issuers. Others preferred the simplicity of centralized systems, and hybrid models blending decentralization with government oversight.
 \begin{figure}[!t]
 \vspace{5pt}
	\centering
	\includegraphics[width=0.6\linewidth]{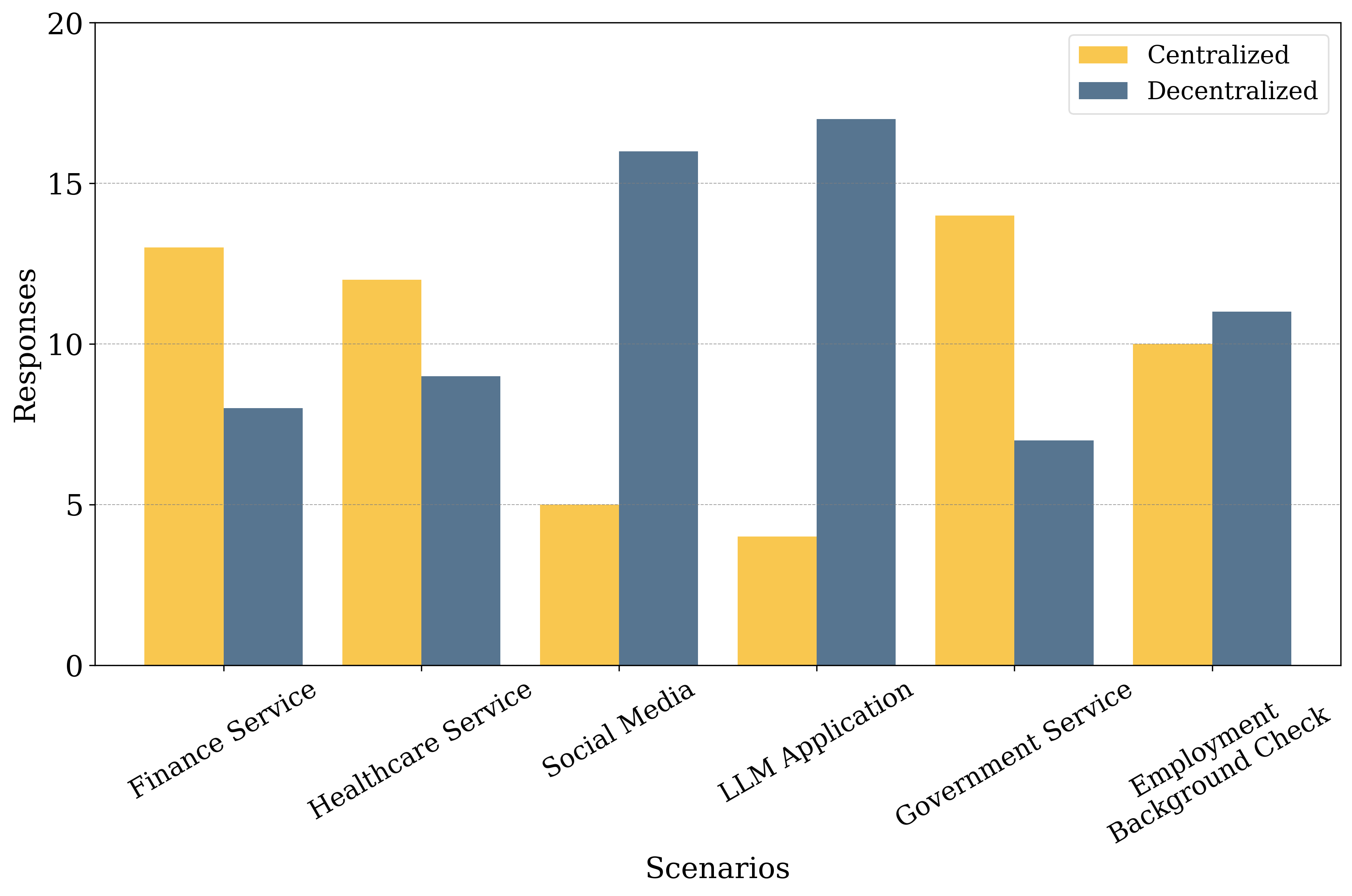}
 \vspace{-15pt}
	\caption{Results of architecture preference: Which type of system would you prefer to issue and manage your PHC for each service provider?
    }
\label{fig:architecture}
\vspace{-15pt}
\end{figure}

\textbf{Decentralized vs. Centralized}
Participants highlighted the risks of centralized storage, preferring decentralization for its enhanced security. P8 stated \textit{"
you've created a wall between any problems with data theft, with a centralized data storage. They only have to breach the walls of one castle with decentralized. They have to breach the walls of two different castles, which makes it a lot more complex and most people won't. Most people won't go through the trouble of doing that. 
"} 
In contrast, participants who expressed a preference for a centralized approach highlighted its simplicity and ease of use. P21 stated \textit{``I feel like central authority can respond to any threat aftermath more effectively for centralize issuance and I guess central, like if gov is issuer then it will have global legitimacy, i can use it for customs.''}

\textbf{In Between Centralization and Decentralization}
Some participants expressed preferences for issuers with a balanced approach of centralized and decentralized systems. 
P1 highlighted a mixed preference, favoring decentralization while emphasizing the government's oversight, \textit{"I think it's sort of in the middle. I think decentralized would be the right way to go, because. as expecting the Government to have enough resources to keep verifying Phcs would be hard...But the verification, like the issuing authority, is still a government, and the Government still has oversight on how these verification systems work."} P3 on the other hand mentioned having decentralized system to issue PHC with having govt as one of the trusted issuers to sign the phc. 

\vspace{-2mm}
\section{RQ3: Design Suggestions for PHC}
\vspace{-2mm}
Participants expressed specific needs for PHC systems, primarily focusing on enhancing security and trust while accommodating diverse needs. From the design sessions, we identified the following themes.

\begin{figure*}[!h]
 \begin{subfigure}{0.48\textwidth}
     \centering
     \includegraphics[width=0.8\textwidth]{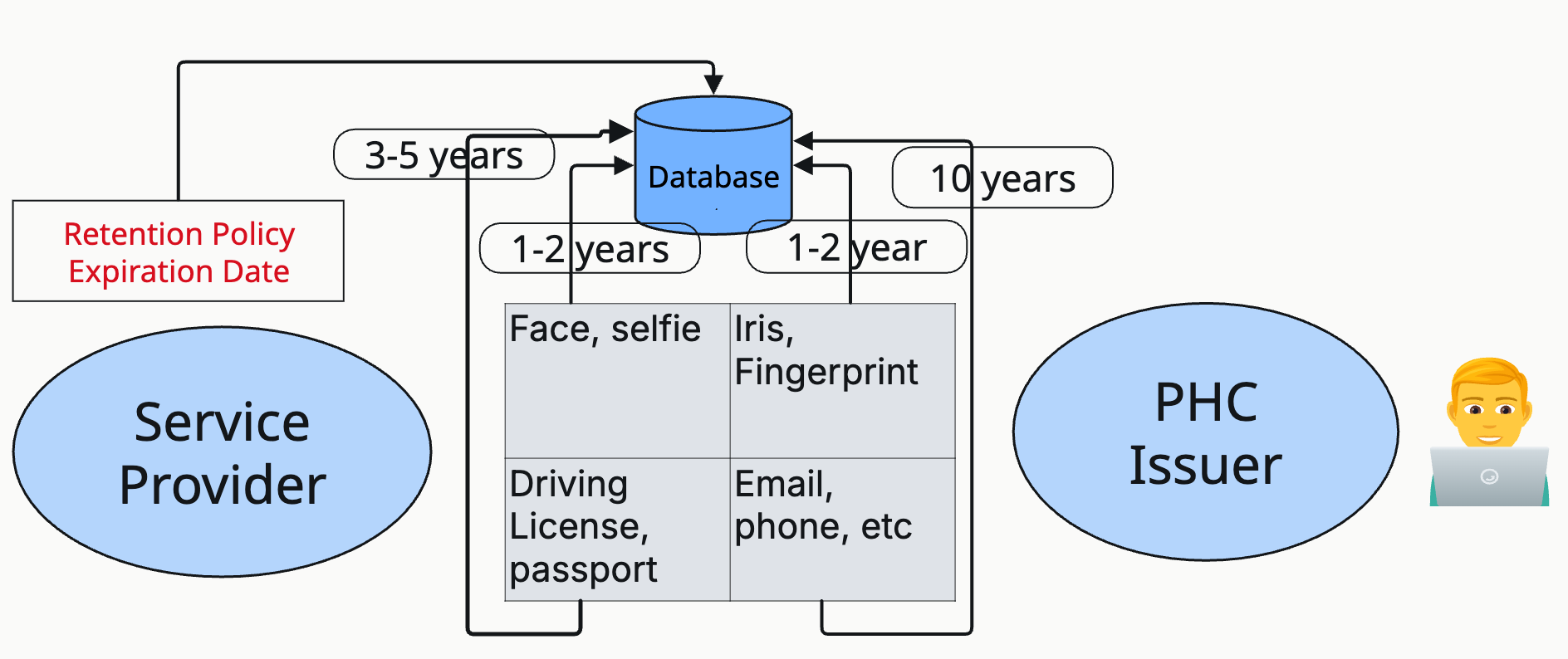}
     \captionsetup{width=\textwidth, font=footnotesize} 
     \caption{Illustration to depict time-bounded credential with retention and expiration date based on different data types}
     \label{fig:P21}
 \end{subfigure}
 \hfill
 \begin{subfigure}{0.48\textwidth}
 \centering
     \includegraphics[width=0.8\textwidth]{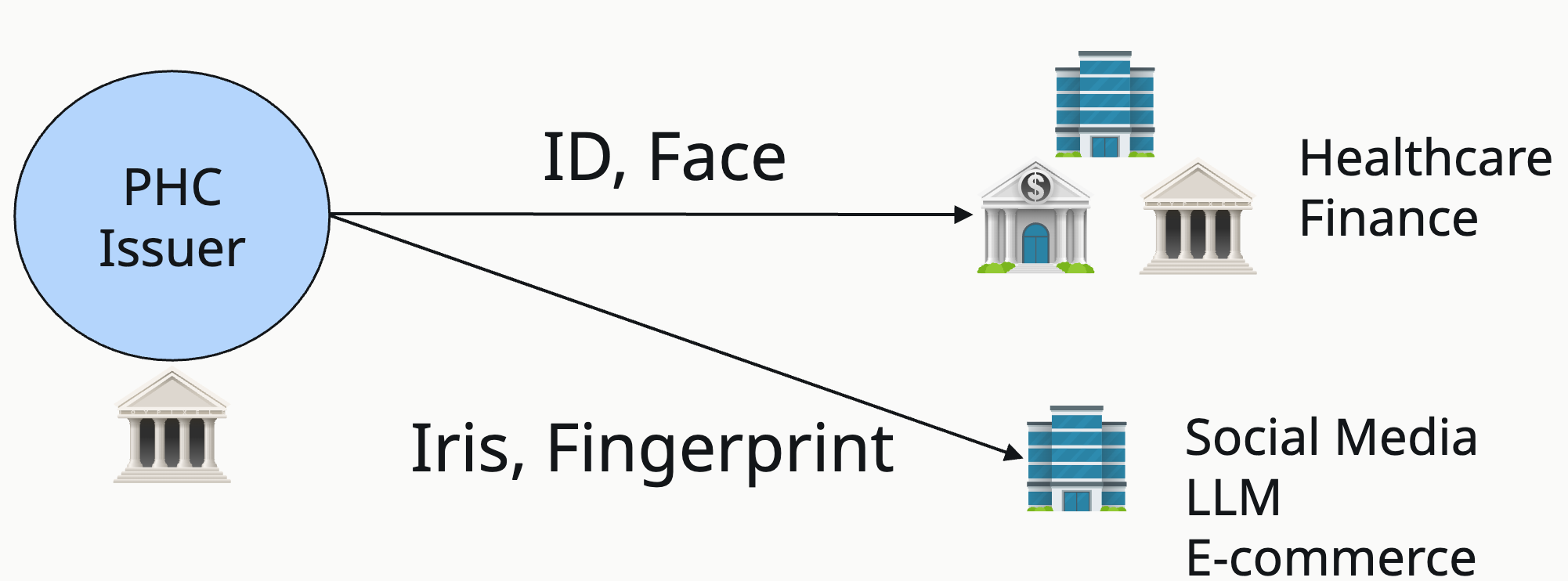}
     \captionsetup{width=\textwidth, font=footnotesize} 
     \caption{Illustration of choice for users to choose various data requirements for PHC issuance for different applications}
     \label{fig:P3}
 \end{subfigure}
 \hfill
 \begin{subfigure}{0.48\textwidth}
 \centering
     \includegraphics[width=0.8\textwidth]{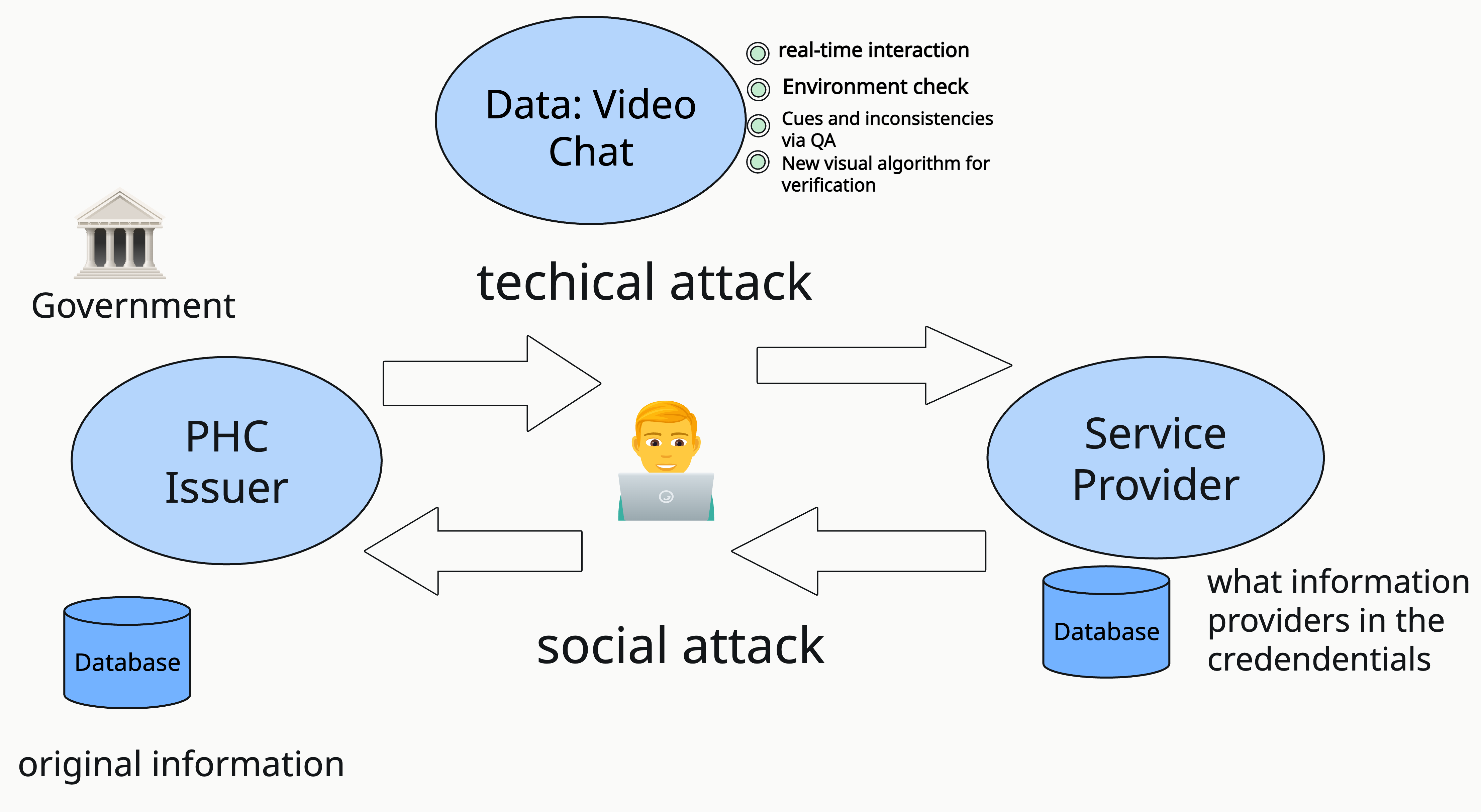}
     \captionsetup{width=\textwidth, font=footnotesize} 
     \caption{Illustration to depict comprehensive visually interactive human check with video chat for humanness cues, environment check}
     \label{fig:P2}
 \end{subfigure}
 \hfill
 \begin{subfigure}{0.48\textwidth}
     \centering
     \includegraphics[width=0.8\textwidth]{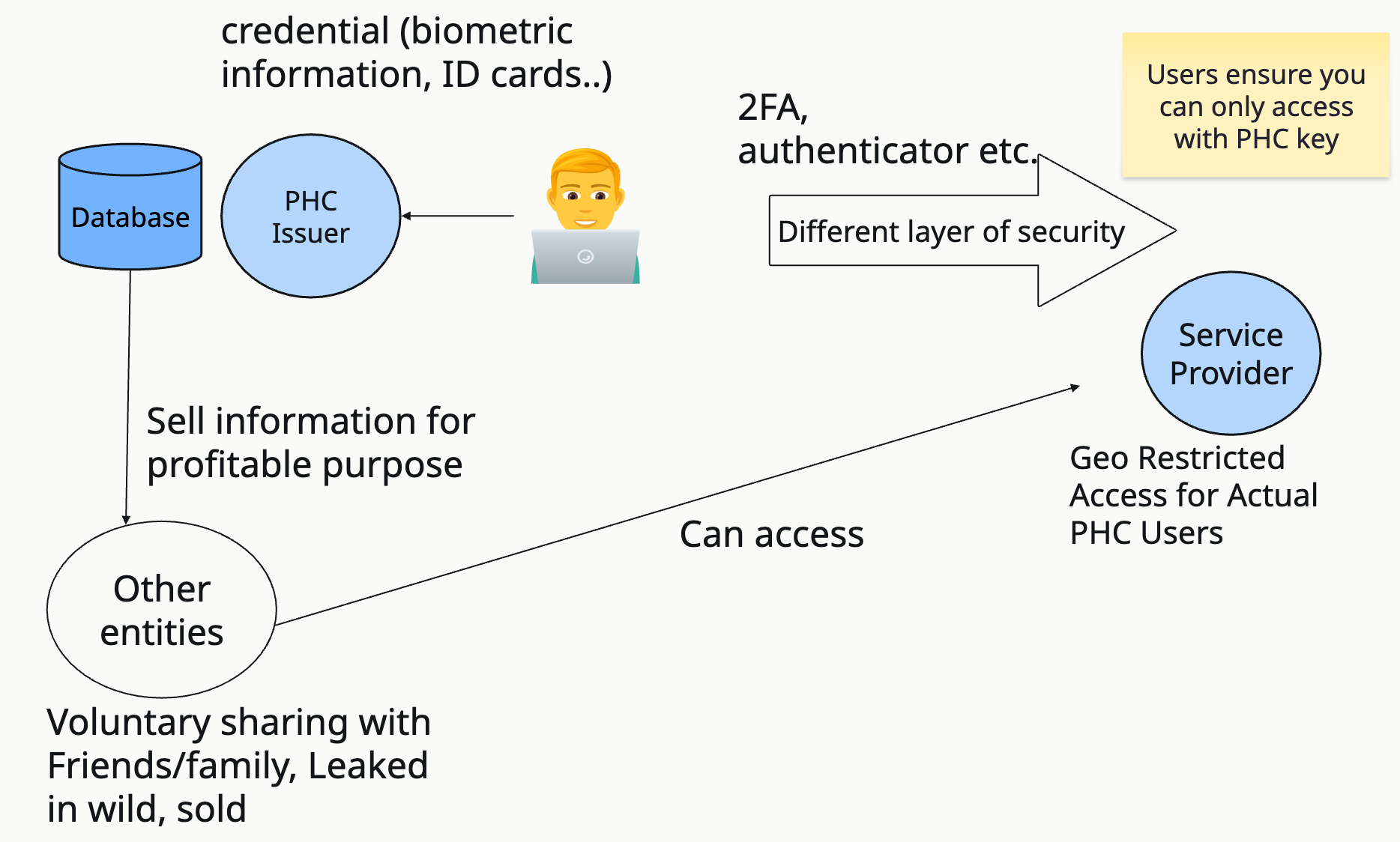}
     \captionsetup{width=\textwidth, font=footnotesize} 
     \caption{Illustration to depict periodic biometrics, dynamic authentication, geo-restricted access}
     \label{fig:P13}
 \end{subfigure}
\hfill
 \begin{subfigure}{0.48\textwidth}
 \centering
     \raisebox{1cm}{
    \includegraphics[width=0.8\textwidth]{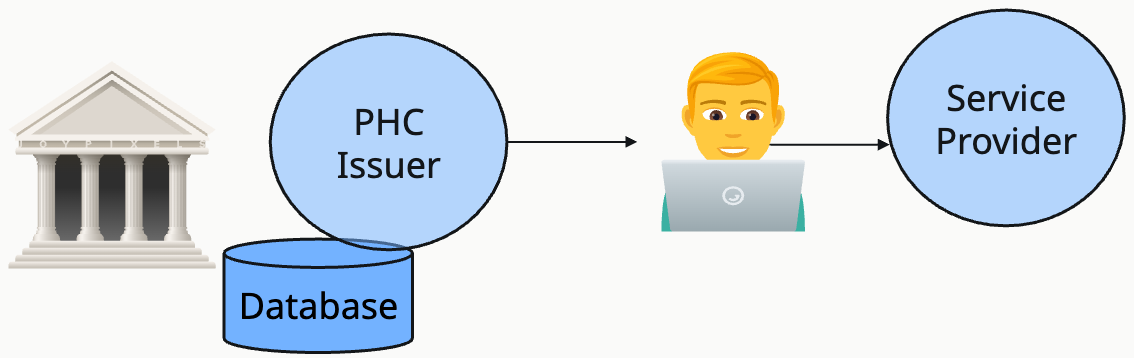}
     }
     \captionsetup{width=\textwidth, font=footnotesize} 
     \caption{Illustration to depict single issuer system with the government}
     \label{fig:P9}
 \end{subfigure}
 \hfill
 \begin{subfigure}{0.48\textwidth}
     \centering\includegraphics[width=0.8\textwidth]{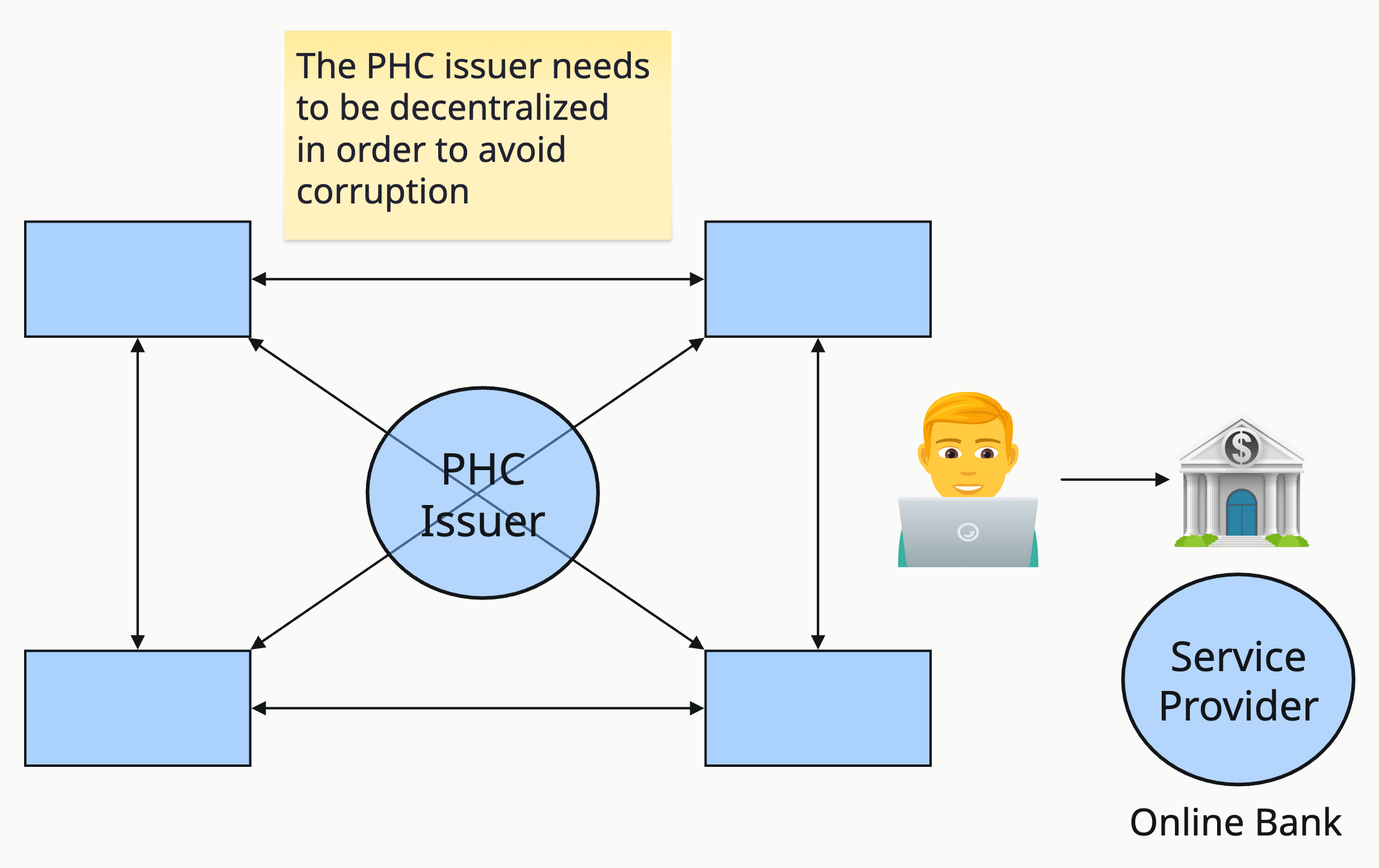}
     \captionsetup{width=\textwidth, font=footnotesize} 
     \caption{Illustration to depict decentralization for PHC governance}
     \label{fig:P7}
 \end{subfigure}
    \caption{Design suggestions for PHC: each represents the following design themes (a) Time-bounded credential for privacy; (b) Sensitivity-based usable credential choice for PHC; (c) Comprehensive visually interactive
human check; (d) Mitigate PHC misuse; (e)(f) Distribute power across issuer: decentralized vs centralized. }
    \label{fig:four_sketches}
\end{figure*}

\textbf{Design Theme 1: Time-bounded Credential for Privacy}
Some users expect credentials to verify personhood with the least amount of personal data, avoiding detailed personal or biometric data collection for a certain period. Another expectation in the same line was the portability of preferred credentials if they can be used across platforms without re-verifying frequently. To address the amount of data and period of such data collection that has been used for issuing credentials, some features or design concepts were -- limited validity period, proof without storage or pseudonymization of the data when storing it. P21 mentioned - \textit{``I want a credential that works like a trusted pass—valid only for a set time (e.g., 30 days, 6 months, or 1 year), with reminders before it expires. It should prove I am a real person without storing my sensitive details or tying them to my real-world identity, like some level of anonymity. Just give me the freedom to be verified without being exposed.'' }, which is depicted in Figure \ref{fig:P21}.

\textbf{Design Theme 2: Sensitivity-based Usable Credential Choice for PHC}
Multiple participants expressed that their preferences are sensitivity-dependent as presented in Figure \ref{fig:P3}. They suggested a design to incorporate choices for end users based on their perceived level of security needs across services in healthcare, finance, social media, etc.
To illustrate the concept, P3, P4 provides a design where she made a choice of government-issued id or face to obtain a PHC for services in healthcare and finance
In her words- \textit{"I think it very much depends on the scenario. For example, places where which are like the most sensitive. Creating a bank account or an account on a government website, or with a healthcare provider. I think, in terms of financial security, or on the government side, my preferred would be a face scan or a video call to verify my face and uploading a doc like my government issued residence permit, or my passport. Whereas in cases which are doesn't have that much of a security concern. So, having an account on ChatGPT, or a social media account, I would be okay with like fingerprint scan or iris scan."} She considers fingerprint and iris are less sensitive since face scan can potentially reveal one's identity, stating \textit{"If I scan my whole face, you can figure out my identity. I think the iris, fingerprint scanning is poses less of a risk, because it's only scanning your iris''}

\textbf{Design Theme 3: Comprehensive Visually Interactive Human Check.}
Another theme comes up where participants explain how \textit{``video chat''} offers several advantages in preventing social attacks during personhood credential issuance by triangulating real-time interaction, visual verification, and environment check. As P2 said \textit{``video chat could be a way to avoid social attacks in PHC issuing. Yes. this way service provider would ask you a lot of questions to make sure you are not controlled by someone else, and they would ask to move my camera, take videos of my whole room to make sure there's no one else beside me. And of course, it's it would. It would not be a 100\% thing, all the other things [only face, ids] are still the old stuff. This could be a new step to developing PHC algorithm.''} P2 further emphasized how human interaction is still a key when stepping into new technological innovation where in the issuance process, trained staff can assess subtle cues and inconsistencies that automated systems might miss, potentially detecting social engineering attempts or ask personalized questions based on the user's responses and behavior, making it difficult for attackers to prepare scripted answers (Figure \ref{fig:P2}).

\textbf{Design Theme 4: Mitigate PHC Misuse: Periodic Biometrics, Dynamic Authentication, and Geo-Restricted Access.}
Participants suggested enhanced key management for the PHC system to address potential misuse, such as identity selling, credential sharing, and unauthorized account setup (Figure \ref{fig:P13}). Suggestions included periodic biometric verification (e.g., facial recognition or fingerprint scans during login) or random re-authentication to ensure that only the credential holder can access the system (P6, P8, P13, P18, P23). P13 highlighted the risk of credential sharing during onboarding, where users may rely on friends or family for setup assistance, potentially leading to unauthorized access. To mitigate this, P13 proposed combining biometrics with dynamic authentication methods, such as time-sensitive push notifications. Additionally, P13 suggested implementing geo-fencing to restrict access from unfamiliar locations or devices, reducing the risk of misuse if credentials are leaked. These suggestions aim to ensure that PHCs are used securely and by the intended user across different services.


\textbf{Design 5: Distribute Power Across Issuer: Decentralized vs Centralized} 
Participants (P5, P9, P10) frequently highlighted the importance of a single issuer for PHC, prioritizing a unified point of trust. For example, P10 suggested the government as the primary issuer for various services, arguing that involving third parties, such as insurers in health contexts, complicates credential management (Figure \ref{fig:P9}). 
Conversely, others (P1, P7, P11) supported a multi-issuer approach, leveraging blockchain-based systems to enable decentralized storage and sector-specific distribution of power, addressing trust and governance concerns in PHC. 
The main design concept focus on ensuring that any misuse of data by one entity would be immediately flagged by others and distributed control among different issuers when included as decentralized issuance systems. P7 illustrated the concept by 
stating \textit{"The important part is, they will have the same information all at once and they are synced together. So someone is uploading their document or info, all these companies are going to have it in sync together, so that if there is any misuse of information is going on here, the other companies would get to know it."}  (Figure \ref{fig:P7})

%% file: Sections/4_Discussion.tex
\vspace{-6mm}
\section{Discussion}
\vspace{-2mm}
Our findings shed light on a wide range of human factors in designing personhood credentials.
In this section, we discuss how the findings can contribute to current state of knowledge in designing user-centered and secure PHC design. 
\vspace{-2mm}
\subsection{Main Findings}
\vspace{-2mm}

\textbf{Nuanced Credential Preference}
Identity management has long been a focal point of user-facing systems, including social media platforms, gaming environments, and collaborative tools, etc~\cite{gorwa2020unpacking, cetinkaya2007verification, sharma2024future, sharma2024unpacking}. We have also observed technological and ideological shifts towards 
decentralized identity- commonly referred as --self-sovereign identity amidst the criticism of large technology companies' data handling practices~\cite{nytimesCambridgeAnalytica}.The most cited case is decentralized (DIDs), with emerging proposed systems, DECO~\cite{zhang2020deco}, Town-Crier~\cite{zhang2016town} -- where users authorizing the release of personal credentials from user devices to websites for proving certain characteristics about themselves. 
While initiatives such as the W3C’s Decentralized Identifier Working Group aim to develop standards~\cite{identityDecentralizedIdentity, w3cccgDecentralizedIdentifiers}, they largely fail to address the technical and usability goals.
Nonetheless, users' preference on managing (e.g. recovery, data handling, trust on issuer ecosystem) verification credentials remains largely unexplored. 

Our work sheds light on trade-offs people consider in onboarding and managing PHC. Our findings highlight some concerns towards PHCs, partly because of “unknown risk” vectors as a new technology compared to traditional verification. Despite these concerns, we find diverse level of adoption preferences
influenced by the “type of data required” for PHC credential issuance and verification as well as personal “security standards” for different services (e.g, finance, health, government
related). 
We also expanded the findings to include nuanced preferences and underlying reasoning that explains why users have certain preferences, extending beyond what existing literature has limited to identifying preferences alone.
For instance, their preference for government-issued IDs is associated with their familiarity with a traditional method of verification. Also, their varied preference for biometrics is backed by subjective perceptions around efficacy and privacy sensitivity. For instance, they often considered facial recognition more resilient verification process than fingerprints.



\textbf{How to Build Trust \& Scale PHCs}
More recently, a cross-industry collaboration \cite{adler2024personhood} resurfaced the conversation of personhood credential 
\cite{borge2017proof, ford2020identity, de2024personhood, sharma2024experts} emphasizing the need of an ecosystem for PHC issuers to facilitate scalability. A fundamental challenge remains the "chicken-and-egg" dilemma: the absence of a broad ecosystem of PHC issuers hinders the adoption of systems that rely on PHCs, and conversely, the lack of such systems makes it difficult to incentivize the establishment of PHC issuers. Our findings highlight an interdependence and the importance of carefully considering both the issuing ecosystem (e.g., centralized versus decentralized models) and the type of issuer (e.g., government entities versus private organizations) as key factors influencing trust and perceptions of security among users to onboard.


Our results highlighted the perceived benefits by users regarding PHCs, which is fairness in representation, aligning with that prior work's discussed proof of personhood is a key in fair online environments, especially when voting or accessing limited resources by eliminating AI-powered manipulations or duplications
\cite{de2024personhood}. 
However, our research also surfaced significant concerns that must be addressed to improve user trust and acceptance. Participants expressed apprehensions about the centralization of PHC issuance, fearing over-concentration of power and control. Ambiguities in regulations surrounding PHCs further compounded these concerns, reflecting broader uncertainties about governance and accountability. Additionally, users highlighted the need to ensure the authenticity of PHCs while addressing risks associated with the misuse of anonymity. These findings resonate with existing literature on privacy and security education, which underscores the importance of user understanding and clear communication in the adoption of new technologies \cite{acquisti2015privacy}. Future research should focus on increasing the explainability and transparency of PHC systems to address these concerns. 

\vspace{-2mm}
\subsection{Design Implications}
\vspace{-2mm}
Drawing upon users' needs and preferences, we suggest actionable design implication for personhood credentials. 

\textbf{Interface Design to Facilitate Verification Choice.}
Our findings shows users' skepticism, partly because of \textit{unknown risk''} vectors of PHC as a new technology. Despite the unknowns, our findings also indicates diverse level of adoption/onboarding preferences towards PHC issuance, such as, type of data requirement to verify themselves which largely depends on their own \textit{``security standard''} developed for different types of services. With the preferences in mind, one possible way to first clearly add list of purpose, operation and policy of PHC leveraging design such as, privacy nutrition label\cite{kelley2009nutrition, li2022understanding} like Apple, to allow users to know the new tech.
As of credential issuance choice, a tiered system of PHCs with varying levels of verification strength
to allow users to choose ground truth data (e.g., gov id, face, video, fingerprint, social graph, etc) based on their security standards. The interface could explicitly add a design of tiered verification options, each corresponding to a different level of security: Level 1 (Low Sensitivity): email or phone verification; Level 2 (Medium Sensitivity): physical id;  Level 3 (High Sensitivity): Biometric fusion (e.g., facial recognition + voice print);  Level 4 (Very High Sensitivity): Multi-factor PHC (e.g., biometrics + social graph verification+physical id), etc. We can implement modular architecture, allowing easy addition or modification of verification methods as technology evolves. 

\textbf{Portability of Personhood Credential.} Our result reveals varied preferences across different services, ranging from financial, health, government to social media. One expectation of users to have a stremlined approach so they don't need to onboard with multiple verification for services they use. A possible way is to design interoperable cryptographically verifiable credentials. Verifiable credential contains claim about the credential holder, issued by a trusted entity, and can be verified without contacting the issuer to prove themselves across various platforms. This is essentially leveraging emerging solutions, such as, Zero-Knowledge Proofs~\cite{tobin2016inevitable} with design principles and standards, such as, W3C verifiable credentials data model, DIF universal resolver~\cite{mazzocca2024survey}, etc. 

\textbf{Dynamic \& Multi-factor Personhood Verification}
One of the repeated concerns in our study whether PHC can be inadvertently shared/used by friends/family, hacked, stolen.  
To address this concern, a potential design implication is to design a dynamic Multi-Factor Personhood Credential (PHC). First, when users verify themselves for the first time, the prompt can combine interactive biometric such as video call, with interaction knowledge question and sharing physical id to compare several ground truth data to issue a robust PHC. Second, to maintain security, a period biometrics verification with time-based trigger system can be designed to prompt for biometrics verification at random intervals or during high risk activities.

\textbf{Decentralized Standards for Industry-Government Issuance System.}
 Our work indicates issuance system and issuers (e.g. govt, private company; decentralized vs centralized) are one of the main factors leading to security and trust perception, thus the broader adoption. In the same line, fundamental limitation of current approaches to PHC creation is their signal failure to address the chicken-and-egg problem in PHC issuance. Without an ecosystem with a broad base of PHC issuers, systems that leverage PHCs will be slow to arise; conversely, without systems that leverage PHCs, it is hard to motivate the creation of PHC issuers. In the United States, for instance, despite industry coalitions pursuing decentralized identity credentials for at least seven years~\cite{mediumDecentralizedIdentity}, digitally signed state IDs are currently available only in California~\cite{caWalletPilot}. 
 To ensure global accessibility in supporting multiple stakeholders as issuers, the system would incorporate cross-chain interoperability protocols like Polkadot or Cosmos and utilize a permissioned blockchain network (e.g., Hyperledger Fabric) to create a distributed ledger for credential issuance and verification. Smart contracts~\cite{sharma2023mixed} would govern the issuance process, ensuring compliance with predefined standards set by both industry and government entities. 

\vspace{-4mm}
\subsection{Limitations}
\vspace{-2mm}

Our interview study has several limitations. Our recruitment of participants resulted in limited diversity in educational backgrounds and age groups. This may restrict the generalizability of our findings, as individuals with a lower age range might exhibit more preferences toward new technology and can have stricter expectations compared to participants with other
educational backgrounds and age range. 

Second participants who didn't have prior knowledge of PHC responded based on an explanation or information video presented during the study. Their responses might differ if they were provided with hands-on interactive PHC system to better convey and understand the concept.
Building on the findings from this study, future work will focus on developing a functional prototype to provide participants with a more immersive and tangible experience.

\vspace{-2mm}
\section{Conclusion}
\vspace{-2mm}
Our study uncovered diverse user perceptions, including trade-offs between traditional verification methods and emerging approaches such as PHCs, as well as dilemmas between physical and digital verification. Furthermore, we highlighted nuanced preferences for each system design dimension: credentials, issuers, and architectures. Additionally, practical PHC functions such as limited credential validity, sensitivity-based selection, interactive human checks, and distributed issuance architectures were identified through design suggestions.
To our knowledge, this is the first user study to focus on PHCs. Our findings extend beyond PHCs, shedding light on key insights for identity verification.

\section{Ethics Considerations}
Our study design and procedures were reviewed and approved by the Institutional Review Board (IRB). We also considered the following ethical aspects:
\textbf{Disclosures}: All collected data (audio transcripts, sketches, and survey results) did not include personally identifiable information and were analyzed anonymously;
\textbf{Experiments with informed consent}: We ensured informed consent , and participants were informed that their participation was voluntary;
\textbf{Deception}: Participants were fully informed of all aspects of interview participation beforehand, clearly stating the scope of the study and its data collection.

\section{Open Science}
 To ensure transparency of this study, all details of our interviews, including the study procedures and survey contents, are available in the following link \url{https://anonymous.4open.science/r/PHC-user-study-14BB/}. We also ensure reproducibility by providing detailed documentation on how to proceed with our interview in Section \ref{sec:method}.

%% file: Sections/Appendix.tex
\section{Appendix: Tables \& Figures}
\clearpage
\onecolumn
\scriptsize{
\begin{longtable}{p{3cm} p{4cm} p{4.5cm} p{4.5cm}}

    \caption{Results of Competitive Analysis.}\\
    \hline
    \textbf{Name of App} & \textbf{Data Workflow} & \textbf{UI/UX Issue} & \textbf{Security/Privacy Issue} \\ \hline
    \endfirsthead
    \hline
    \textbf{Name of App} & \textbf{Required data} & \textbf{UI/UX Issue} & \textbf{Security/Privacy Issue} \\ \hline
    \endhead
    \hline
    \endfoot
    \hline
    \endlastfoot
        \textbf{CAPTCHA}  
        & Solve CAPTCHA challenge \ding{221} JavaScript tracks user behavior (mouse, keystroke dynamics)  \ding{221} Submit response, which includes collected IP address, device, and browser information.
        & \ding{202} Screen readers don't help visually impaired users to solve graphical CAPTCHAs. \ding{203} It creates a barrier for users with other impairments (motor, cognitive). \ding{204} Sometimes CAPTCHA challenges are so difficult that they make users leave the website or application.
        & Users are often unaware of the data collected by CAPTCHA, which could potentially be used beyond verification.\\
        
        \textbf{reCAPTCHA}  
        & Click checkbox \ding{221} JavaScript tracks user behavior (mouse, keystroke dynamics, touchscreen interactions)  \ding{221} Submit response, which includes collected IP address, cookies, device, and browser information.
        & If the system regards users as bots, there is no way they can access it.
        & Users are often unaware of the data collected by reCAPTCHA, which could potentially be used beyond verification.\\

        \textbf{Idena}  
        & Create password \ding{221} Import private key backup to sign in \ding{221} Solve the flip challenge on the scheduled validation ceremony date.
        & The flip tests with time constraints make passing difficult, especially for users with cognitive or motor impairments. 
        & \ding{202} While flip challenges are not directly tied to personal information, they could potentially reveal patterns of the user. \ding{203} If a user's private key is compromised, an attacker can gain full control over the account.\\

        \textbf{World ID}  
        & Enter birth date (Android only) \ding{221} Enter phone number (OTP) \ding{221} Enter username and password \ding{221} Connect Google Drive or iCloud \ding{221} Scan iris in-person at the Orb location.
        & \ding{202} The onboarding screen is overly simplified, lacking sufficient explanations for users to understand the system's workflow and how their data is managed. \ding{203} The backup process requiring Google Drive access may cause confusion about unintended access to personal space.
        & Collecting sensitive biometric data, such as iris scans, requires high transparency and accountability about its handling and protection.\\

        \textbf{Humanode}  
        & Generate an asymmetric key pair \ding{221} Insert private key \ding{221}  Scan face to link biometric data with public key.
        & Users may struggle with complex operations like node setup and validation, leading to confusion and onboarding friction.
        & Collecting high-precision biometric data, such as 3d mapping face scans, requires high transparency and accountability about its handling and protection.\\

        \textbf{BrightID}  
        & Enter name, photo, and password \ding{221} Scan connections' QR code.
        & Users may find it confusing to navigate the verification process (Aura, Bitu verification).
        & Although the data is not stored centrally, there are still privacy concerns regarding connections and the information that is publicly visible to others.\\

        \textbf{Proof of Humanity}  
        & Connect wallet \ding{221} Submit name, front-facing photo, a video showing face, wallet address on a physical surface, and stating the required phrase.
        & Users may not be aware of the risk of losing their deposit if their submission fails.
        & \ding{202} Requiring users' photos and videos raises privacy concerns, leading to potential trust issues with PoH issuers. \ding{203} The wallet address used to submit a profile will be publicly linked to user's identity. The user's wallet holdings and transaction history will be publicly linked. \\

        \textbf{DECO}  
        & Extracts encrypted session data of the website the user accessed \ding{221} Generate zero-knowledge proof for specific information \ding{221} Verifier validates the proof without accessing the original source.
        & The advanced technical nature, relying on mechanisms such as zero-knowledge proofs and TLS, makes it difficult for users to understand how their information is processed and protected.
        & \ding{202} Improperly implemented ZKPs may unintentionally leak sensitive information. \ding{203} It allows proving a substring of the response, but this could be exploited through context manipulation.\\

        \textbf{CANDID}  
        & Hold a digital credential \ding{221} Select attributes for disclosure \ding{221} Generate a zero-knowledge proof \ding{221} Verifier validates the proof without requiring full credentials.
        & The advanced technical nature, relying on mechanisms such as zero-knowledge proofs and decentralized identity, makes it difficult for users to understand how their information is processed and protected.
        & Repeated use of certain identity attributes across different services could lead to deanonymization. \\

        \textbf{India Aadhaar Card}  
        & Apply Aadhaar with proof of identity, proof of address, proof of date of birth, and proof of relationship documents \ding{221} Scan fingerprints, iris, and face at an enrollment center.
        & Users must submit a combination of documents, which may not be accessible for those in rural areas with limited access to government services.
        & \ding{202} Data breaches have been reported, with millions of Indians' Aadhaar and passport details compromised and exposed on the dark web. \ding{203} Collecting multiple biometrics raises privacy concerns, as compromised biometric data is permanent and cannot be reset like passwords.\\

        \textbf{Estonia e-ID}  
        & Enter email address \ding{221} Apply e-ID with CV, credit/debit card information, photo of face, and passport or EU ID card.
        & The reliance on external card readers makes it less accessible for non tech-savvy users who have difficulty installing and configuring the card reader.
        & \ding{202} The e-ID system serves as a single point of access for multiple e-services, which could be problematic if compromised. \ding{203} In 2017, a security flaw was discovered in the chip encryption, potentially allowing identity theft and unauthorized access to private data.\\

        \textbf{Japan My Number Card}  
        & Provide one photo ID or two non-photo IDs with an issue notice at a government office.
        & \ding{202} Users must go through lengthy procedures to obtain and renew the card, often requiring in-person visits. \ding{203} It requires four passwords (three 4-digit PINs and one 6-16 character alphanumeric password), with resets only available at a government office.
        & \ding{202} My Number is printed on the back of the card, posing a privacy vulnerability when physically handed to others. \ding{203} The system's expansion to cover health insurance has raised concerns about an increased risk of unauthorized cross-referencing of personal information.\\

        \textbf{China’s Social Credit System}  
        & The system collects and integrates four categories data: basic information, information on administrative penalties and permits, any irregularities, and red list or blacklist information (if applicable).
        & \ding{202} Users often don't understand how their scores are calculated or what specific actions impact their ratings. \ding{203} With multiple government and private systems in place, China's social credit system lacks unified standards.
        & The systems gather vast amounts of personal information for economic reliability and social behavior, raising surveillance concerns. \\

        \textbf{ID.me}  
        & Enter email and password \ding{221} Chose MFA options (SMS etc.) \ding{221} Take a photo of government-issued ID.
        & Users without a government-issued ID cannot verify their identity, excluding those who lack access to such documents.
        & Sensitive personal information is centrally stored, making users' credentials highly vulnerable if compromised.\\

        \textbf{Civic Pass}
        & Connect wallet \ding{221} Verify via video selfie or CAPTCHA or government-issued ID.
        & Users with multiple wallets struggle with uniqueness verification because the system associates one face with one wallet.
        & On-chain attestation could expose metadata that links a user’s identity to blockchain transactions, potentially compromising privacy.\\
       

\label{tab:litcomparison}
\end{longtable}
}